\documentclass{article}
\usepackage{amssymb,color,graphicx}
\usepackage{amssymb}
\usepackage{wrapfig,framed,cancel}
\usepackage[colorlinks=true, pdfstartview=FitV, linkcolor=black, citecolor=black, urlcolor=blue]{hyperref}
\usepackage{bm}
\usepackage[thinlines]{easymat}
\definecolor{shadecolor}{rgb}{0.93, 0.93, 0.86}
\begingroup\makeatletter\ifx\SetFigFont\undefined%
\gdef\SetFigFont#1#2#3#4#5{%
\reset@font\fontsize{#1}{#2pt}%
\fontfamily{#3}\fontseries{#4}\fontshape{#5}%
 \selectfont}%
\fi\endgroup%

\newtheorem{theorem}{Theorem}[section]
\newtheorem{prop}{Proposition}[section]

\newtheorem{cor}{Corollary}[section]

\newtheorem{remark}{Remark}[section]
\newtheorem{problem}{Problem}[section]

\def\br{\begin{remark}}
\def\er{\end{remark}}

\def\C{{\mathbb C}}

\def\bN{{\mathbb N}}
\def\bR{{\mathbb R}}
\def\eqref#1{ (\ref{#1})}
\def\&{&\hspace{-20pt}}
\def\R{{\mathbb R}}
\def\d{{\rm d}}

\def\1{\mathbf 1}

\def \pa{\partial}

\def\bea{\begin{eqnarray}}
\def\eea{\end{eqnarray}}
\def\Id{\mathrm{Id}}
\def\Ai{\mathrm{Ai}}
\def\f{{\bm f}}
\def\g{{\bm g}}

\def\Cc{{\mathcal C}}
\def\F{{\mathcal F}}
\def\G{{\mathcal G}}

\def\0{{\bf 0}}

\def\res{\mathop{{\rm res}}}

\def\Ai{\mathrm{Ai}}
\def\T{\mathrm{T}}
\def\B{{\mathcal{B}}}
\def\hB{{\hat\mathcal{B}}}

\def\A{\mathcal A}
\def\U{\mathcal U}

\def\QED{ {\bf Q.E.D}\par \vskip 4pt}
\def\K{{\mathbb{K}}}
\newcommand{\be}{\begin{eqnarray}}
\newcommand{\ee}{\end{eqnarray}}
\newcommand{\bes}{\begin{eqnarray*}}
\newcommand{\ees}{\end{eqnarray*}}
\newcommand{\ds}{\displaystyle}

\def\I{{\mathcal I}}
\def\KF{{\mathbb{F}}}
\def\KG{{\mathbb{G}}}

\definecolor{light-blue}{rgb}{0.8,0.85,1}
\definecolor{blue}{rgb}{0,0,1}
\definecolor{red}{rgb}{1,0,0}

\def\le{\left}
\def\ri{\right}
\def\ba{\begin{eqnarray}}
\def\eeq{\end{eqnarray}}

\renewcommand{\theequation}{\arabic{section}.\arabic{equation}}
\makeatletter
\@addtoreset{equation}{section}
\makeatother

\title{Non-commutative Painlev\'e equations and \\ Hermite-type matrix orthogonal polynomials \footnote{The work of the first author is partially supported by the ANR grant DIADEMS, the work of the  second author is partially supported
by MTM2012-36732-C03-03 (Ministerio de Economia y Competitividad), FQM-262, FQM-4643, FQM-7276 (Junta de Andalucia) and Feder Funds (European Union). We are also grateful to the Research Group on Orthogonal Polynomials and Approximation Theory of the University of Sevilla and to the G\'eanpyl project, for financing visits to Sevilla and Angers.}}
\author{Mattia Cafasso$^{\dagger}$ and  Manuel D. de la Iglesia$^{\ddagger}$ \\
   \footnotesize $\dagger$ \footnotesize LUNAM Universit\'e, LAREMA, Universit\'e d'Angers\\
    \footnotesize 2 Boulevard Lavoisier, 49100 Angers, France. cafasso@math.univ-angers.fr \\
   $\ddagger$  \footnotesize
    \  Departamento de An\'{a}lisis Matem\'{a}tico.
   Universidad de Sevilla \\
   \footnotesize Apdo (P. O. BOX) 1160. 41080 Sevilla, Spain.
mdi29@us.es
\\
\ \ }
\date{}

\begin{document}

\maketitle

\begin{abstract}
We study double integral representations of Christoffel-Darboux kernels associated with two examples of Hermite-type matrix orthogonal polynomials. We show that the Fredholm determinants connected with these kernels are related through the Its-Izergin-Korepin-Slavnov (IIKS) theory with a certain Riemann-Hilbert problem. Using this Riemann-Hilbert problem we obtain a Lax pair whose compatibility conditions lead to a non-commutative version of the Painlev\'e IV differential equation for each family.
\end{abstract}

\section{Introduction}

Let us denote with $(p_n)_{n\in\bN}$ the classical Hermite polynomials such that $\mathrm{deg}(p_n) = n$ and
$$
\int_\bR p_n(x)p_m(x){\rm e}^{-x^2}dx = \delta_{nm}.
$$
As we know from the pioneering work of Gaudin and Mehta \cite{MehtaBook}, the so-called Hermite kernel
$$
K_n(x,y) := \sum_{k=0}^{n-1} p_k(x)p_k(y){\rm e}^{-\frac{x^2+y^2}2},
$$
describes the statistical properties of the eigenvalues of a random matrix in the space of $(n\times n)$ Hermitian matrices equipped with the measure $\mu(M):={\rm e}^{-\mathrm{Tr}(M^2)}dM$, where $dM$ denotes the standard Haar measure. More precisely, $\mu(M)$ induces a measure on the space of configurations of $n$ points on the real line, hence a determinantal point process whose particles are given by the eigenvalues of $M$, and whose correlation functions $\rho_k(x_1,\ldots,x_k)$ are given by the formula
$$\rho_k(x_1,\ldots,x_k) = \det(K_n(x_i,x_j))_{i,j=1}^k.$$
In particular, the \emph{last particle distribution} $F(s)$, describing the probability that the largest eigenvalue is smaller than $s$, is given by the Fredholm determinant $F(s) = \det(\Id-\chi_s\K_n),$ where we denoted with $\chi_s$ the indicator function of the semi infinite interval $[s,\infty)$, and $\K_n$ is the integral operator whose kernel is $K_n(x,y)$ (for a very nice introduction about determinantal random point processes and random matrices see \cite{JohanssonReview}). A remarkable connection between $F(s)$ and the Painlev\'e IV equation has been discovered in the nineties by Tracy and Widom. Namely, in \cite{TWPIV}, they proved that the log-derivative $R(s):=\partial_s\log(F(s))$ solves the sigma-form of Painlev\'e IV equation
$$(R'')^2 + 4(R')^2 (R' + 2n) - 4 (sR' - R)^2 = 0.$$
The aim of this article is to extend this result to the case of Christoffel-Darboux kernels associated to Hermite-type \emph{matrix-valued} orthogonal polynomials (MOP). To that purpose we will show first how to obtain double integral representations of some examples of MOP.

First, we need to introduce some notations. In what follows we consider $N$ fixed and we use preferably boldface letters to denote matrices, and standard font for scalars.
We  also use $\bm I_N$ for the $(N\times N)$ identity matrix, omitting the explicit reference to its dimension when it cannot lead anyone into confusion. Let $\mathcal{C}$ be any piecewise smooth oriented curve. A \emph{weight matrix} $\bm W=(W_{ij})_{i,j = 1}^N:\mathcal{C}\rightarrow GL(N,\mathbb{R})$
on the curve $\mathcal{C}$ is a positive definite matrix at any point of the curve with finite moments. We say that a matrix function $\bm F$ belongs to the space $L^2_{\bm W}(\mathcal{C},\mathbb{R}^{N\times N})$ if
$$
    \int_{\mathcal{C}}\bm F(z)\bm W(z)\bm F^\T(z) dz<\infty,
$$
where the superscript ``T'' denotes, as usual, the transpose\footnote{More generally one could consider matrix-valued complex weights, and in this case you have to substitute the transpose with the Hermitian conjugate.}. In the above definition we mean that the integral is finite entry by entry. In the case when the weight matrix $\bm W$ is the identity matrix $\bm I_N$ we will just write $L^2\left(\mathcal{C},\mathbb{R}^{N\times N}\right)$. This induces a matrix-valued inner product for any two matrix functions $\bm F, \bm G \in L^2_{\bm W}(\mathcal{C},\mathbb{R}^{N\times N})$, denoted by
\begin{equation}\label{Innerp}
    \langle\bm F,\bm G\rangle_{\bm W}=\int_{\mathcal{C}} \bm F(z)\bm W(z)\bm G^{\T}(z)dz.
\end{equation}
This is not an inner product in the common sense, but it has properties similar to the usual scalar inner products. It is also possible to define a scalar norm of a matrix function $\bm F$ by $\mbox{Tr}\left(\langle\bm F,\bm F\rangle_{\bm W}\right)^{1/2}$ (see \cite{DPS}). Therefore $L^2_{\bm W}(\mathcal{C},\mathbb{R}^{N\times N})$ with this norm is a Hilbert space and \eqref{Innerp} is the inner product.

A sequence $(\bm P_n)_{n\in\bN} $ of orthonormal MOP with respect to a weight matrix $\bm W$ is a sequence of matrix polynomials satisfying
$$
\deg \bm{P}_n = n,\quad \langle \bm{P}_n,\bm{P}_m\rangle_{\bm W}=\bm I_N\delta_{nm},\quad\forall\; n,m\in\bN.
$$

We will mainly work with MOP on the real line, therefore $\mathcal{C}=\mathbb{R}$. Work in the last few years has revealed a number of explicit families of MOP on the real line. In many cases they are joint eigenfunctions of some fixed differential operator with matrix coefficients independent of the degree $n$ of the polynomials. This study was initiated in \cite{D1}, but nontrivial examples had to wait until \cite{DG1, GPT1}. These examples are the matrix analogue of the classical families of Hermite, Laguerre and Jacobi polynomials.

Given a complete orthonormal family of MOP $(\bm P_n)_{n\in\bN} $ in $L^2_{\bm W}(\mathbb{R},\mathbb{R}^{N\times N})$ the Christoffel-Darboux (CD) kernel is defined as (see, for instance, (2.26) of \cite{GdIM})
\begin{equation}\label{MVHKI}
    \bm K_n(x,y):=\sum_{k=0}^{n-1}\bm P_k^\T(y)\bm P_k(x),\quad x,y\in\mathbb{R}.
\end{equation}
We observe immediately the following properties:
\begin{enumerate}
  \item $\bm K_n(x,y)=\bm K_n^\T(y,x)$.
  \item $\forall\; \bm F\in L^2_{\bm W}\left(\mathbb{R},\mathbb{R}^{N\times N}\right),\quad \bm F(y)=\langle 	\bm F(x),\bm K_n(x,y)\rangle_{\bm W}\quad \mathrm{(reproducing\; kernel\; property)}.
  $
  \item $\bm K_n(x,z)=\langle \bm K_n^\T(z,y),\bm K_n(y,x)\rangle_{\bm W}$.
\end{enumerate}
Observe, in particular, from the second equation, that we are thinking about the kernel as an integral operator acting \emph{on the left}\footnote{It is possible to work with a CD kernel acting \emph{on the right}, but in that case we have to consider a different inner product defined by $\left(\bm F,\bm G\right)_{\bm W}:=\int_{\mathbb{R}}\bm G^\T(x)\bm W(x)\bm F(x)dx.$ Now the CD kernel will be defined by $\bm K_n^\T(x,y):=\sum_{k=0}^{n-1}\bm P_k^\T(x)\bm P_k(y) = \bm K_n(y,x)$.} for functions in $L^2_{\bm W}\left(\mathbb{R},\mathbb{R}^{N\times N}\right)$.

In Section \ref{SEC2} we find double integral representations of the CD kernel \eqref{MVHKI} (or rather a slight modification of \eqref{MVHKI}) for two Hermite-type families of MOP, already introduced in \cite{DG1}. We first find integral representations of the families $(\bm P_n)_{n\in\bN}$ using the corresponding second-order differential equation that they satisfy. To the best of our knowledge, this is the first time that integral representations of families of MOP are studied in detail.

Secondly, in Section \ref{SEC3}, we study the Fredholm determinant $\det(\Id-\chi_s\mathbb{K}_n)$ of the integral operator $\mathbb{K}_n$ with kernel $\bm K_n(x,y)$. Our main tool will be the theory of integrable operators \`a la Its-Izergin-Korepin-Slavnov; see \cite{IIKS} and also \cite{DeiftIntOp} for a survey of the remarkable properties of these operators and applications to statistical mechanics, random matrices and orthogonal polynomials. More specifically, the Fredholm determinant $\det(\Id-\chi_s\mathbb{K}_n)$ will be identified with the isomonodromic tau function associated to a specific Riemann-Hilbert problem (see Appendix \ref{AP1} and \ref{AP2}). This Riemann-Hilbert problem, through a standard procedure, can be reduced to one with constant jumps, leading to a certain Lax system of equations. The compatibility conditions of the Lax system give a couple of matrix ordinary differential equations. Combining these two equations we obtain a non-commutative version of the derived nonlinear Painlev\'e IV differential equation for each family, see Theorems \ref{theoncPIVfirstcase} and \ref{theoncPIVsecondcase}. The contents of this section are very close, in spirit, to the one obtained by one of the authors and Marco Bertola in \cite{BC2}.

Finally, in Section \ref{3.1.1}, we give a symmetric formulation of the non-commutative Painlev\'e IV equation. The non-commutative Painlev\'e II equation used in \cite{BC2} has been introduced by Retakh and Rubtsov in \cite{RetakhRubtsov}, where the authors obtained this equation as a reduction of a non-commutative analogue of Toda equations. Hence it appears desirable to verify if also the non-commutative Painlev\'e IV equations obtained in this article can be written, analogously, as reductions of some suitable non-commutative Toda-type equations. The content in Section \ref{3.1.1} goes in this direction.



\section{Hermite-type MOP and related CD kernels}\label{SEC2}

In this section we consider a couple of examples of MOP already introduced in \cite{DG1}. Both are orthogonal with respect to Hermite-type weight matrices of the form
$$
    \bm W(x):={\rm e}^{-x^2}\bm T(x)\bm T^\T(x),\quad x\in\mathbb{R},
$$
where $\bm T$ is certain matrix polynomial. Consider an orthonormal family of MOP $(\bm P_n)_{n\in\bN}$. Now define, for every $n\in\bN$, the orthonormal function
$$
\bm\Phi_n(x):={\rm e}^{-x^2/2}\bm P_n(x)\bm T(x).
$$
$(\bm\Phi_n)_{n\in\bN}$ is a family of matrix functions orthonormal with respect to the identity matrix, i.e.
$$
\langle \bm\Phi_n,\bm\Phi_m\rangle_{\bm I_N}=\bm I_N\delta_{nm},\quad \forall\; n,m\in\bN,
$$
where the inner product $\langle \cdot,\cdot\rangle$ is defined by \eqref{Innerp}. In our examples the family $(\bm\Phi_n)_{n\in\bN}$ will always be complete in the space $L^2\left(\mathbb{R},\mathbb{R}^{N\times N}\right)$ (see Section 6 of \cite{dI} for more details).

The Hermite-type CD kernel is then defined, slightly modifying \eqref{MVHKI}, as\footnote{The kernel defined in this section, indeed, is equal to the one defined in the introduction up to a conjugation by ${\rm e}^{-\frac{x^2}2}T(x)$. In order to keep the notation simple we used, nevertheless, the same symbol.}
\begin{equation}\label{MVHK}
    \bm K_n(x,y):=\sum_{k=0}^{n-1}\bm\Phi_k^\T(y)\bm\Phi_k(x).
\end{equation}

In order to find double integral representations of these two examples of matrix-valued kernels \eqref{MVHK} we will use the fact that the corresponding MOP are eigenfunctions of a second-order differential equation of the form
\begin{equation}\label{sode}
    \bm P_n''(x)\bm F_2(x)+\bm P_n'(x)\bm F_1(x)+\bm P_n(x)\bm F_0=\bm\Gamma_n\bm P_n(x),
\end{equation}
where $\bm F_2, \bm F_1$ and $\bm F_0$ are matrix polynomials (which do not depend on $n$) of degrees less than or equal to 2, 1 and 0, respectively, and the eigenvalue $\bm\Gamma_n$ is a symmetric matrix.

\subsection{The first example}\label{SEC21}

Let $\bm A_N$ be the $(N\times N)$ nilpotent  ($\bm A^N_N=\bm 0)$ matrix
\begin{equation}\label{AAA}
\bm A_N:=\sum_{i=1}^{N-1}\nu_i\bm E_{i,i+1},\quad \nu_i\in\mathbb{R},
\end{equation}
where $\bm E_{i,j} = (\delta_{ir}\delta_{sj})_{r,s=1}^N$ is the elementary matrix with 1 at entry $(i,j)$ and 0
elsewhere, and $\bm J_N$ the diagonal matrix
\begin{equation}\label{JJ}
\bm J_N:=\sum_{i=1}^N(N-i)\bm E_{i,i}.
\end{equation}
Again, we will sometimes remove the dependence of $N$ and write $\bm A=\bm A_N$ and $\bm J=\bm J_N$, whenever there is no confusion about the dimension of the matrices. $\bm A$ and $\bm J$ satisfy the algebraic relation $[\bm A,\bm J]=-\bm A$.

Let $\bm W$ be the following weight matrix
\begin{equation}\label{W1}
    \bm W(x)={\rm e}^{-x^2}{\rm e}^{\bm Ax}{\rm e}^{\bm A^\T x},\quad x\in\mathbb{R},
\end{equation}
already introduced in \cite{DG1}. Observe that ${\rm e}^{\bm Ax}$ is an upper triangular matrix polynomial of degree $N-1$ (since $\bm A^N = \bm 0$).

The family of MOP
\begin{equation}\label{FAM1}
\bm P_n(x)={\rm e}^{-\bm A^2/4}\widehat{\bm P}_n(x),
\end{equation}
where $(\widehat{\bm P}_n)_{n\in\bN}$ denotes the monic orthogonal family with respect to \eqref{W1}, satisfies a second-order differential equation as in \eqref{sode} (see, for instance, Section 4 of \cite{dI}) where
$$
\bm F_2(x)=\bm I,\quad \bm F_1(x)=-2x\bm I+2\bm A,\quad \bm F_0(x)=\bm A^2-2\bm J,\quad \bm \Gamma_n=-2n\bm I-2\bm J.
$$
The family \eqref{FAM1} is not orthonormal, but it will be normalized later for the computation of the CD kernel \eqref{MVHK}. In the following, given a matrix $\bm M$, we will use the standard notation $z^{\bm M}:= {\rm e}^{\bm M\log z}$, and the branch cut of $\log z$ is chosen to be the real negative axis.

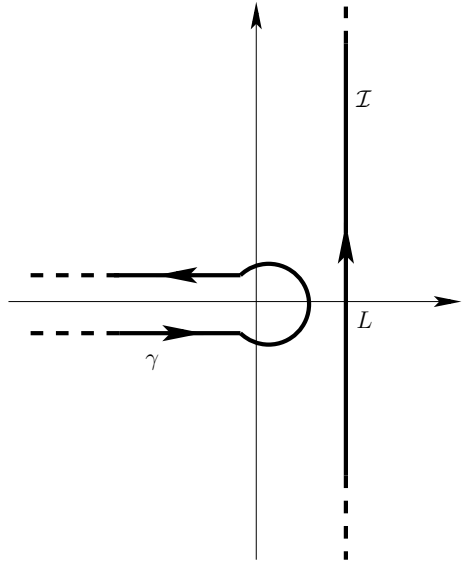
\begin{wrapfigure}{r}{0.35\textwidth}
\resizebox{0.4\textwidth}{!}{\begin{picture}(0,0)%
\includegraphics{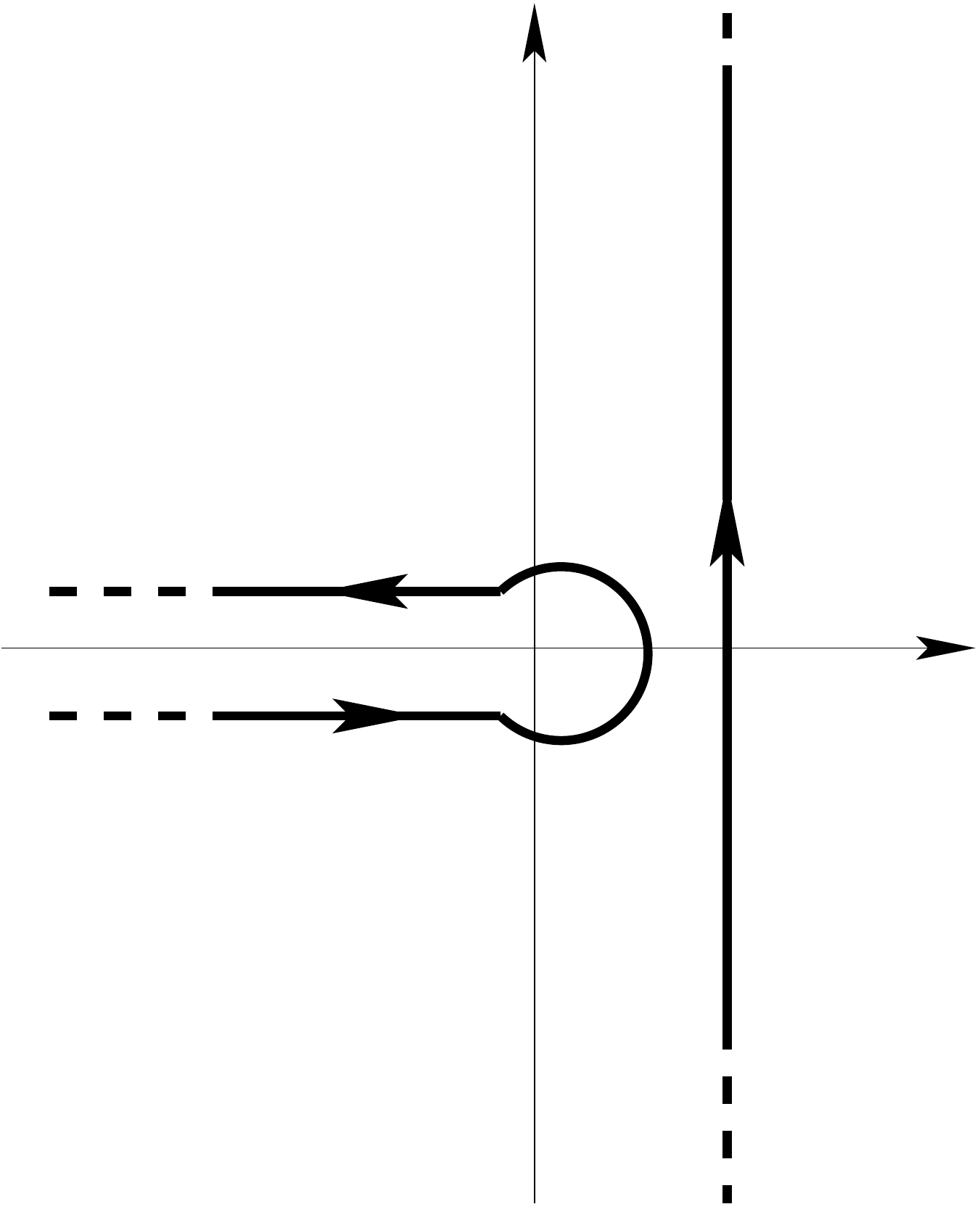}%
\end{picture}%
\setlength{\unitlength}{3947sp}%
%
\begin{picture}(6474,8016)(2839,-8365)
\put(7801,-1861){\makebox(0,0)[lb]{\smash{{\SetFigFont{20}{24.0}{\rmdefault}{\mddefault}{\updefault}{\color[rgb]{0,0,0}$\I$}%
}}}}
\put(4801,-5536){\makebox(0,0)[lb]{\smash{{\SetFigFont{20}{24.0}{\rmdefault}{\mddefault}{\updefault}{\color[rgb]{0,0,0}$\gamma$}%
}}}}
\put(7801,-5011){\makebox(0,0)[lb]{\smash{{\SetFigFont{20}{24.0}{\rmdefault}{\mddefault}{\updefault}{\color[rgb]{0,0,0}$L$}%
}}}}
\end{picture}%
}
\caption{Contours $\gamma$ and $\I$.}
\label{contours}
\end{wrapfigure}

\begin{theorem}\label{Thm1}
Let $(\bm P_n)_{n\in\bN}$ be the family of MOP defined by \eqref{FAM1}. Then there exist suitable constant matrices $\bm C_n$ and $\bm D_n$ such that
\begin{equation}\label{IntRep1}
    \bm P_n(x) {\rm e}^{\bm Ax}=\oint_{\gamma} z^{-\bm J}\bm C_n z^{\bm J} {\rm e}^{-z^2+2zx}\frac{dz}{z^{n+1}},
\end{equation}
and
\begin{equation}\label{IntRep2}
    \bm P_n(x) {\rm e}^{\bm Ax}= {\rm e}^{x^2}\int_{\mathcal{I}} w^{\bm J}\bm D_n w^{-\bm J}{\rm e}^{w^2-2xw}w^ndw,
\end{equation}
where the contour $\gamma$ encloses the origin, closes at $-\infty$ and it is traversed in a counterclockwise direction while $\mathcal{I}:=L+i\mathbb{R}$, and $L>0$ is chosen so to have no intersection between $\gamma$ and $\I$ (see Figure \ref{contours}).
\end{theorem}
\noindent\emph{Proof:} For the first integral representation \eqref{IntRep1} we observe, using the formula ${\rm e}^{-\bm Ax}\bm Je^{\bm Ax}=\bm J+x\bm A$, that the functions $\bm Y_n(x)=\bm P_n(x)e^{\bm Ax}$ satisfy the following differential equation
\begin{equation}\label{FDEq}
    \bm Y_n''(x)-2x\bm Y_n'(x)-2\bm Y_n(x)\bm J+2(n\bm I+\bm J)\bm Y_n(x)=\bm 0.
\end{equation}
Let us look for solutions of (\ref{FDEq}) of the form
$$
\bm Y_n(x)=\int_{\eta}\bm V_n(z) {\rm e}^{2zx}dz,
$$
where $\eta$ is some contour in the $z$-plane. Substituting this expression into the differential equation \eqref{FDEq} and integrating by parts, it is easy to see that \eqref{FDEq} holds if the two following conditions are satisfied
\begin{equation}\label{fir}
\bm V_n'(z)=-\left(2z+\frac{n+1}{z}\right)\bm V_n(z)+\frac{1}{z}\left[\bm V_n(z)\bm J-\bm J\bm V_n(z)\right],
\end{equation}
\be\label{bound}
z\bm V_n(z){\rm e}^{2zx}\bigg|_{\eta}=\bm 0.
\ee

General solutions of homogeneous first-order matrix differential
equations of the form
$$
\bm X'(z)=\bm A(z)\bm X(z)+\bm X(z)\bm B(z)
$$
are easy to obtain considering solutions $\bm X(z)=\bm X_1(z)\bm C\bm X_2(z)$, where $\bm C$ is any constant matrix, such that $\bm X_1'(z)=\bm A(z)\bm X_1(z)$ and $\bm X_2'(z)=\bm X_2(z)\bm B(z)$ (see for instance \cite{W}, Lemma 30.1). Applying this to the equation (\ref{fir}), there exist, for every $n\geq0$, constant matrices $\bm C_n$ such that the general solution is
of the form $\bm V_n(z)=\bm V_{n,1}(z)\bm C_n\bm V_{n,2}(z)$, where
$$
\bm V_{n,1}'(z)=-\bigg(2z\bm I+\frac{n+1}{z}\bm I+\frac{\bm J}{z}\bigg)\bm V_{n,1}(z),\quad \bm V_{n,2}'(z)=\bm V_{n,2}(z)\frac{\bm J}{z}.
$$
Therefore
$$
\bm V_n(z)=z^{-\bm J}\bm C_n z^{\bm J}\frac{{\rm e}^{-z^2}}{z^{n+1}}.
$$
gives a solution for \eqref{fir}. Now we observe that, choosing $\eta = \gamma$ (the same contour as in the case of scalar Hermite polynomials), the expression on the left hand side of \eqref{bound} is nothing but the residue at infinity of the expression $z\bm V_n(z) {\rm e}^{2xz}$, which is clearly zero because of the term $ {\rm e}^{-z^2}$. Hence
$$
    \bm Y_n(x):=\oint_{\gamma} z^{-\bm J}\bm C_n z^{\bm J} {\rm e}^{-z^2+2zx}\frac{dz}{z^{n+1}}
$$
is a solution of \eqref{FDEq}. In particular, since $\bm P_n(x) {\rm e}^{\bm Ax}$ satisfies a differential equation as in \eqref{FDEq}, there will be suitable constant matrices $\bm C_n$, which depend on the family $(\bm P_n)_{n\in\bN}$, such that we get the integral representation \eqref{IntRep1}.\\

For the second integral representation \eqref{IntRep2} we follow a similar argument, just observing that the functions $\bm Z_n(x)= {\rm e}^{-x^2}\bm P_n(x) {\rm e}^{\bm Ax}$ satisfy the differential equation
\begin{equation}\label{FDEq2}
    \bm Z_n''(x)+2x\bm Z_n'(x)-2\bm Z_n(x)\bm J+2((n+1)\bm I+\bm J)\bm Z_n(x)=\bm 0,
\end{equation}
and considering Fourier type solutions of \eqref{FDEq2} of the form
$$
\bm Z_n(x)=\int_{-\infty}^{\infty}\bm V_n(t){\rm e}^{2ixt}dt.
$$
Then (as before) we find some first-order matrix differential equation for $\bm V_n(z)$ and we conclude that
$$
\bm V_n(t)=t^{\bm J}\tilde{\bm C}_n t^{-\bm J} {\rm e}^{-t^2}t^{n},
$$
where $\tilde{\bm C}_n, n\geq0,$ are certain constant matrices. Hence
$$
    \bm Z_n(x)=\int_{-\infty}^{\infty} t^{\bm J}\tilde{\bm C}_n t^{-\bm J} {\rm e}^{-t^2+2ixt}t^ndt.
$$
In particular, since $ {\rm e}^{-x^2}\bm P_n(x) {\rm e}^{\bm Ax}$ satisfies a differential equation as in \eqref{FDEq2}, by the change of variables $t\rightarrow iw$, followed by a shift $iw\rightarrow iw+L$, we conclude that there will be suitable constant matrices $\bm D_n=i^{n+1}i^{\bm J}\tilde{\bm C}_n i^{-\bm J}$, depending on the family $(\bm P_n)_{n\in\bN}$, such that \eqref{IntRep2} holds.
\QED

\begin{remark}
Observe that the constant matrices $\bm C_n$ and $\bm D_n$ obtained initially from the differential equations \eqref{FDEq} and \eqref{FDEq2}, respectively, have many degrees of freedom. It is exactly the choice of the family $(\bm P_n)_{n\in\bN}$ which allows us to have a unique representation of $\bm C_n$ and $\bm D_n$, as we will see below. The meaning of this is because there are many families of matrix polynomials satisfying \eqref{FDEq} or \eqref{FDEq2}, but not all of them are orthogonal with respect to some weight matrix.
\end{remark}

%

\begin{cor}
The Hermite-type CD kernel \eqref{MVHK} associated with the MOP with weight \eqref{W1} can be written as
\begin{equation}\label{MVHKg}
     \bm K_n(x,y)={\rm e}^{(x^2-y^2)/2}\int_{\mathcal{I}}dw\oint_{\gamma}dz \bm M_n(z,w){\rm e}^{w^2-2xw-z^2+2zy},
\end{equation}
where
\begin{equation}\label{suma}
     \bm M_n(z,w):=z^{\bm J}\left[\sum_{k=0}^{n-1}\bm C_k^{\T}z^{-\bm J}\|\bm P_k\|_{\bm W}^{-2}w^{\bm J}\bm D_k\left(\frac{w}{z}\right)^k\right]\frac{1}{z}w^{-\bm J},
\end{equation}
and $\|\bm P_k\|_{\bm W}^{2}$ denotes the matrix-valued norm with respect to \eqref{Innerp}.
\end{cor}
\noindent\emph{Proof:}
It is just enough to use the formula
$$\bm\Phi_n(x)={\rm e}^{-x^2/2}\|\bm P_n\|_{\bm W}^{-1}\bm P_n(x){\rm e}^{\bm Ax}$$ and then the integral representations \eqref{IntRep1} (with the substitution $x\rightarrow y$) and \eqref{IntRep2} in the definition of the matrix-valued kernel \eqref{MVHK}.
\QED

The coefficients $\bm C_n$ and $\bm D_n$ are not easy to obtain for our example since there are no general structural formulas like the norms or the coefficients of our family of MOP for any general size $N$. We will study in detail the case $N=2$ in the next subsection.

\subsubsection{A detailed study of the case $N=2$}\label{firstcase}

Let us consider the $(2\times 2)$ case related to MOP with respect to the weight
$$
\bm W(x) =  {\rm e}^{-x^2} {\rm e}^{\bm A_2x} {\rm e}^{\bm A_2^\T x},\quad \bm A_2 := \left( \begin{array}{cc}
														0 & \nu\\
														0 & 0
													\end{array}\right),\quad\nu\in\R.
$$
We have many structural formulas for this example (see \cite{DG2}). In particular, the polynomials $(\bm P_n)_{n\in\bN}$ in \eqref{FAM1} have diagonal norms
\begin{equation}\label{norms1}
\|\bm P_n\|_{\bm W}^{2}=\frac{n!\sqrt{\pi}}{2^n}\left(
                                          \begin{array}{cc}
                                            \gamma_{n+1}^2 & 0 \\
                                          0 & 1/\gamma_{n}^2 \\
                                          \end{array}
                                        \right),
\end{equation}
 where we denoted
$$
    \gamma_{n}^2:=1+\frac{n}{2}\nu^2.
$$

The coefficients $\bm C_n$ and $\bm D_n$ of the integral representations \eqref{IntRep1} and \eqref{IntRep2} are given by
\begin{equation}\label{Cn1}
\bm C_n=\frac{n!}{2^{n+1}\pi i}\left(
                                          \begin{array}{cc}
                                            1 & \nu(n+1)/2 \\
                                          -\nu/\gamma_{n}^2 & 1/\gamma_{n}^2 \\
                                          \end{array}
                                        \right),
\end{equation}
and
\begin{equation}\label{Dn1}
\bm D_n=\frac{1}{i\sqrt{\pi}}\left(
                                          \begin{array}{cc}
                                            1 & \nu\\
                                          -\displaystyle\frac{n\nu}{2\gamma_{n}^2} & 1/\gamma_{n}^2 \\
                                          \end{array}
                                        \right).
\end{equation}
These coefficients are computed directly from the integral representations \eqref{IntRep1} and \eqref{IntRep2}, since we have that the family $(\bm P_n)_{n\in\bN}$ can be given in terms of scalar Hermite polynomials (see \cite{DG2}). Therefore \eqref{Cn1} is obtained from \eqref{IntRep1} using Cauchy's residue theorem and \eqref{Dn1} is obtained from \eqref{IntRep2} using standard manipulations of the Fourier transform of Hermite polynomials.

\begin{prop}
The Hermite-type CD kernel \eqref{MVHKg}, for $N=2$, can be written as
\begin{equation}\label{MVHKR}
\bm K_n(x,y)=\frac{2}{(2\pi i)^2}{\rm e}^{(x^2-y^2)/2}\int_{\mathcal{I}} dw\oint_{\gamma} dz \; z^{\bm J_2} \bm B_nz^{-\bm J_2}w^{\bm J_2}\bm B_n^{-1} w^{-\bm J_2}\ds\frac{ {\rm e}^{w^2-2xw-z^2+2zy+n\log(w/z)}}{w-z},
\end{equation}
where
\begin{equation}\label{Bn1}
\bm B_n :=   \left( \begin{array}{cc}
1 &-\nu\\
          				\ds\frac{n\nu}{2} & 1
\end{array}\right),\quad \det(\bm B_n)=\gamma_{n}^2.
\end{equation}
\end{prop}
\noindent\emph{Proof:}
Computing \eqref{suma}, using \eqref{norms1}, \eqref{Cn1} and \eqref{Dn1}, we have that
$$
\bm M_n(z,w)=\frac{2}{(2\pi i)^2}\left(
                                          \begin{array}{cc}
                                          -\ds\frac{1}{w-z}+\ds\frac{\left(\ds\frac{w}{z}\right)^{n}}{w-z} \left[\ds\frac{z(\gamma_{n}^2-1)+w}{w\gamma_{n}^2}\right]
                                           & \nu\left(\ds\frac{\left(\ds\frac{w}{z}\right)^{n}}{\gamma_{n}^2}-1\right)\\
                                           \\
                                          \ds\frac{n\nu}{2\gamma_{n}^2}\ds\frac{\left(\ds\frac{w}{z}\right)^{n}}{zw} & -\ds\frac{1}{w-z}+\ds\frac{\left(\ds\frac{w}{z}\right)^{n}}{w-z} \left[\ds\frac{w(\gamma_{n}^2-1)+z}{z\gamma_{n}^2}\right]\\
                                          \end{array}
                                        \right).
$$
Then, using Cauchy's residue theorem, we observe that the diagonal terms $-1/(w-z)$ and the $-\nu$ in the upper right corner of $\bm M_n(z,w)$ do not give any contribution. Hence we have
$$
\bm K_n(x,y)=\frac{2}{(2\pi i)^2} {\rm e}^{(x^2-y^2)/2}\int_{\mathcal{I}} dw\oint_{\gamma} dz \; \tilde{\bm M}_n(z,w)\ds\frac{ {\rm e}^{w^2-2xw-z^2+2zy+n\log(w/z)}}{w-z}
$$
where
$$\tilde{\bm M}_n(z,w) =
\left(\begin{array}{cc}
                  \ds\frac{z(\gamma_{n}^2-1)+w}{w\gamma_{n}^2} & \ds\frac{\nu}{\gamma_{n}^2}(w-z)\\
                  \\
                               \ds\frac{n\nu}{2\gamma_n^2}(z^{-1}-w^{-1}) & \ds\frac{w(\gamma_{n}^2-1)+z}{z\gamma_{n}^2} \end{array}\right),$$
and finally it is easy to see that
$$
\tilde{\bm M}_n(z,w) = z^{\bm J_2} \bm B_nz^{-\bm J_2}w^{\bm J_2}\bm B_n^{-1} w^{-\bm J_2},
$$
where $\bm B_n$ is given by \eqref{Bn1}.
\QED
	
%

\begin{remark}
Also for the case $N=3$, the Hermite-type CD kernel \eqref{MVHKg} has a representation analogue to \eqref{MVHKR}, for a given $(3\times 3)$ constant matrix $\bm B_n$ (we do not report here the computations, since they are completely analogues but more cumbersome). We conjecture that for any $N\geq 1$ and a given $n\geq0$, there always exists a constant matrix $\bm B_n$ such that
\be\label{MVHK1c}
\bm K_n(x,y)\!\!=\!\!\frac{2}{(2\pi i)^2} {\rm e}^{(x^2-y^2)/2}\!\!\int_{\mathcal{I}}\! dw\oint_{\gamma}\! dz \; z^{\bm J_N} \bm B_nz^{-\bm J_N}w^{\bm J_N}\bm B_n^{-1} w^{-\bm J_N}\ds\frac{{\rm e}^{w^2-2xw-z^2+2zy+n\log(w/z)}}{w-z}.
\ee
\end{remark}

\subsection{The second example}\label{SEC22}

Let $\bm W$ be the following weight matrix
\begin{equation}\label{W2}
    \bm W(x)= {\rm e}^{-x^2} {\rm e}^{\bm Bx^2} {\rm e}^{\bm B^\T x^2},\quad x\in\mathbb{R},
\end{equation}
where $\bm B=\bm A(\bm I+\bm A)^{-1}$ and $\bm A$ is the $(N\times N)$ nilpotent matrix \eqref{AAA}. The family of MOP
\begin{equation}\label{FAM2}
\bm P_n(x)=[(\bm I+\bm A)^{-1/2}]^{2n+1}\widehat{\bm P}_n(x),\quad n\in\bN,
\end{equation}
where $(\widehat{\bm P}_n)_{n\in\bN}$ denotes the monic orthogonal family with respect to \eqref{W2}, satisfies a second-order differential equation as in \eqref{sode} (see Section 5 of \cite{dI}) with
$$
\bm F_2(x)=\bm I,\quad \bm F_1(x)=2x(2\bm B-\bm I),\quad \bm F_0(x)=2(\bm B-2\bm J),\quad \bm\Gamma_n=-2n\bm I-4\bm J,
$$
where $\bm J$ is the diagonal matrix \eqref{JJ}. Also in this case, in complete analogy with the first example, we can deduce two different integrable representations of the polynomials $(\bm P_n)_{n\in\bN}$ and, consequently, an integral representation of the related Hermite-type CD kernel.

\begin{theorem}
Let $(\bm P_n)_{n\in\bN}$ be the family of MOP defined by \eqref{FAM2}. Then there exist suitable constant matrices $\bm C_n$ and $\bm D_n$ such that
\begin{equation}\label{IntRep12}
    \bm P_n(x) {\rm e}^{\bm Bx^2}=\oint_{\gamma} z^{-2\bm J}\bm C_n z^{2\bm J} {\rm e}^{-z^2+2zx}\frac{dz}{z^{n+1}},
\end{equation}
and
\begin{equation}\label{IntRep22}
    \bm P_n(x) {\rm e}^{\bm Bx^2}= {\rm e}^{x^2}\int_{\mathcal{I}} w^{2\bm J}\bm D_n w^{-2\bm J}{\rm e}^{w^2-2xw}w^ndw,
\end{equation}
where the contour $\gamma$ encloses the origin, closes at $-\infty$ and it is traversed in a counterclockwise direction while $\mathcal{I}:=L+i\mathbb{R}$, and $L>0$ is chosen so to have no intersection between $\gamma$ and $\I$ (see again Figure \ref{contours}).
\end{theorem}
\noindent\emph{Proof:}
This theorem is proven exactly in the same way as Theorem \ref{Thm1}, using as starting point the formula $ {\rm e}^{-\bm Bx^2}\bm J {\rm e}^{\bm Bx^2}=\bm J+x^2(\bm B-\bm B^2)$ instead of ${\rm e}^{-\bm Ax}\bm J{\rm e}^{\bm Ax} = \bm J+x\bm A$.
\QED

\begin{cor}
The Hermite-type CD kernel \eqref{MVHK} can be written as
\begin{equation}\label{MVHKg2}
     \bm K_n(x,y)= {\rm e}^{(x^2-y^2)/2}\int_{\mathcal{I}}dw\oint_{\gamma}dz \bm M_n(z,w) {\rm e}^{w^2-2xw-z^2+2zy},
\end{equation}
where
\begin{equation}\label{suma2}
     \bm M_n(z,w)=z^{2\bm J}\left[\sum_{k=0}^{n-1}\bm C_k^{\T}z^{-2\bm J}\|\bm P_k\|_{\bm W}^{-2}w^{2\bm J}\bm D_k\left(\frac{w}{z}\right)^k\right]\frac{1}{z}w^{-2\bm J}.
\end{equation}
and $\|\bm P_k\|_{\bm W}^{2}$ denotes the matrix-valued norm with respect to \eqref{Innerp}.
\end{cor}

As before, the coefficients $\bm C_n$ and $\bm D_n$ are not easy to obtain in general for this example. We will focus on the case $N=2$.

\subsubsection{A detailed study of the case $N=2$}\label{secondcase}
Let us consider the $(2\times 2)$ case related to matrix-valued polynomials orthogonal with respect to the weight
$$
\bm W(x)= {\rm e}^{-x^2} {\rm e}^{\bm B_2x^2} {\rm e}^{\bm B_2^\T x^2},\quad \bm B_2=\bm A_2.
$$
Again, from \cite{DG2}, we have many structural formulas for this example. In particular, the polynomials $(\bm P_n)_{n\in\bN}$ in \eqref{FAM2} have diagonal norms
\begin{equation}\label{norms2}
\|\bm P_n\|_{\bm W}^{2}=\frac{n!\sqrt{\pi}}{2^n}\left(
                                          \begin{array}{cc}
                                            \delta_{n+2}^2 & 0 \\
                                          0 & 1/\delta_{n}^2 \\
                                          \end{array}
                                        \right),
\end{equation}
where we denoted
\begin{equation}\label{gam2}
    \delta_{n}^2:=1+\frac{n(n-1)}{4}\nu^2.
\end{equation}
The coefficients $\bm C_n$ and $\bm D_n$ of the integral representations \eqref{IntRep12} and \eqref{IntRep22} are given by
\begin{equation}\label{Cn2}
\bm C_n=\frac{n!}{2^{n+1}\pi i}\left(
                                          \begin{array}{cc}
                                            1 & \displaystyle\frac{\nu(n+1)(n+2)}{4} \\
                                          -\nu/\delta_{n}^2 & 1/\delta_{n}^2 \\
                                          \end{array}
                                        \right),
\end{equation}
and
\begin{equation}\label{Dn2}
\bm D_n=\frac{1}{i\sqrt{\pi}}\left(
                                          \begin{array}{cc}
                                            1 & \nu\\
                                          -\displaystyle\frac{n(n-1)\nu}{4\delta_{n}^2} & 1/\delta_{n}^2 \\
                                          \end{array}
                                        \right),
\end{equation}
which can be computed in a similar way as the first example.

\begin{prop}
The Hermite-type CD kernel \eqref{MVHKg2}, for $N=2$, can be written as
\begin{equation}\label{MVHKR2}
\bm K_n(x,y)=\frac{2}{(2\pi i)^2}{\rm e}^{(x^2-y^2)/2}\int_{\mathcal{I}} dw\oint_{\gamma} dz \; z^{2\bm J_2} \bm B_nz^{-\bm J_3}w^{\bm J_3}\hat{\bm B}_n w^{-2\bm J_2}\ds\frac{{\rm e}^{w^2-2xw-z^2+2zy+n\log(w/z)}}{w-z},
\end{equation}
where $\bm B_n$ is a $(2 \times 3)$ matrix and $\hat{\bm B}_n$ a $(3 \times 2)$ matrix given by
\begin{equation}\label{Bn2}
\bm B_n :=   \left( \begin{array}{ccc}
\ds\frac{1}{\delta_{n+1}^2}& \ds\frac{n\nu^2}{2\delta_{n+1}^2\delta_{n}^2} &-\nu\\
\\
          				\ds\frac{\nu n(n+1)}{4\delta_{n+1}^2} & -\ds\frac{n\nu}{2\delta_{n+1}^2\delta_{n}^2} & 1
\end{array}\right),\quad \hat{\bm B}_n :=   \left( \begin{array}{cc}
1 &\nu\\
\\
1&\nu\\
\\
\ds-\frac{\nu n(n-1)}{4\delta_{n}^2} & \ds\frac{1}{\delta_{n}^2}
\end{array}\right).
\end{equation}
$\hat{\bm B}_n$ is a right inverse of $\bm B_n$, i.e. $\bm B_n\hat{\bm B}_n=\bm I_2$.
\end{prop}
\noindent\emph{Proof:}
Computing \eqref{suma2}, using \eqref{norms2}, \eqref{Cn2} and \eqref{Dn2}, we have that
$$
\bm M_n(z,w)=\frac{2}{(2\pi i)^2}\left(\begin{array}{cc}
                                            \ds\frac{\left(\ds\frac{w}{z}\right)^{n-2}-1}{w-z}+\ds\frac{\left(\ds\frac{w}{z}\right)^{n-1}}{w\delta_{n}^2}+ \ds\frac{\left(\ds\frac{w}{z}\right)^{n}}{w\delta_{n+1}^2}& \nu\left(-w-z+\ds\frac{z\left(\ds\frac{w}{z}\right)^{n}}{\delta_{n}^2}+\ds\frac{w\left(\ds\frac{w}{z}\right)^{n}}{\delta_{n+1}^2}\right)\\
                                            \\
                                          \nu\left(\ds\frac{n(n+1)\left(\ds\frac{w}{z}\right)^{n}}{4\delta_{n+1}^2z^2w}+\ds\frac{n(n-1)\left(\ds\frac{w}{z}\right)^{n}}{4\delta_{n}^2zw^2}\right) & \ds\frac{\left(\ds\frac{w}{z}\right)^{n+2}-1}{w-z}-\ds\frac{\left(\ds\frac{w}{z}\right)^{n+1}}{z\delta_{n+1}^2}- \ds\frac{\left(\ds\frac{w}{z}\right)^{n}}{z\delta_{n}^2}\\
                                          \end{array}
                                        \right).
$$
Then, using Cauchy's residue theorem, we observe that the diagonal terms $-1/(w-z)$ and the $-w-z$ in the upper right corner of $\bm M_n(z,w)$ do not give any contribution. Now taking out the term $\le(\frac{w}{z}\ri)^{n}(w-z)^{-1}$ of the above expression we observe that it can be written as $z^{2\bm J_2}\bm B_nz^{-\bm J_3}w^{\bm J_3}\hat{\bm B}_n w^{-2\bm J_2}$ where $\bm B_n$ and $\hat{\bm B}_n$ are given by \eqref{Bn2}. The equality $\bm B_n\hat{\bm B}_n=\bm I_2$ is a consequence of the definition of $\delta_{n}^2$, see \eqref{gam2}.
\QED
	
\begin{remark}
As before, we also analyzed the case $N=3$. The Hermite-type CD kernel has a similar representation \eqref{MVHKR2} for certain constant matrices $\bm B_n$ and $\hat{\bm B}_n$ of size $(3 \times 5)$ and $(5 \times 3)$, respectively, satisfying $\bm B_n\hat{\bm B}_n=\bm I_3$. We conjecture that for any $N\geq 1$ and a given $n\geq0$, there always exist constant matrices $\bm B_n$ and $\hat{\bm B}_n$ of size $(N\times(2N-1))$ and $((2N-1)\times N)$, respectively, satisfying $\bm B_n\hat{\bm B}_n=\bm I_N$ and such that
\be\label{MVHK2c}
\bm K_n(x,y)\!\!=\!\!\frac{{\rm e}^{(x^2-y^2)/2}}{2(\pi i)^2} \!\!\int_{\mathcal{I}}\!\!dw\oint_{\gamma}\!\!dz \; z^{2\bm J_N}\bm B_nz^{-\bm J_{2N-1}}w^{\bm J_{2N-1}}\hat{\bm B}_n w^{-2\bm J_N}\ds\frac{ {\rm e}^{w^2-2xw-z^2+2zy+n\log(w/z)}}{w-z}.
\ee
\end{remark}

\section{Hermite-type kernels and non-commutative Painlev\'{e} IV}\label{SEC3}

Let us consider a $(N\times N)$ matrix-valued kernel of the general form
\be\label{kernel}
  \bm K_n(x,y)=\frac{2}{(2\pi i)^2}{\rm e}^{(x^2-y^2)/2}\int_{\mathcal{I}}dw\oint_{\gamma}dz \;
    \bm\B_n(z)\bm\hB_n(w)\ds\frac{{\rm e}^{w^2-2xw-z^2+2zy+n\log(w/z)}}{w-z},
\ee
where $\bm\B_n$ and $\bm\hB_n$ are, respectively, an $(N\times p)$ and a $(p\times N)$ square-integrable matrix function such that $\bm\B_n(z)\bm\hB_n(z) = \bm I_N$. We think of $\bm K_n(x,y)$ as a kernel of an integral operator $\K_n$ acting \emph{on the left} for matrix functions, namely
$$
[\K_n \bm F](x) := \int_\R \bm F(y)\bm K_n(x,y)dy\quad\forall \; \bm F\in L^2(\R,\R^{N\times N}).
$$
The Hermite-type CD kernels studied in Section \ref{SEC2} are particular cases of this one.

In the following $\chi_s$ will always denotes the indicator function of the interval $[s,\infty)$. We start with the following theorem:

\begin{theorem}\label{TH1}
	Consider, given $s\in\R$, the operator $\tilde \K_{n,s}: L^2(\gamma \cup \I,\mathbb{R}^{N\times N}) \longrightarrow L^2(\gamma \cup \I,\mathbb{R}^{N\times N}) $ with kernel (acting on the right)
	$$\tilde{\bm K}_{n,s}(w,z) := \frac{{\rm e}^{-\frac{z^2}2 + 2s(z-w) -n\log(z)}}{2\pi i (w-z)}\chi_\I(w)\chi_{\gamma}(z)+
	\frac{{\rm e}^{-\frac{w^2}2+z^2+n\log(z)}\bm\B(z)\bm\hB(w)}{2 \pi i (w-z)}\chi_{\gamma}(w)\chi_{\I}(z).$$
	The following equality between Fredholm determinants holds:
$$
		\det(\Id-\chi_{s}\K_{n}) = \det(\Id-\tilde \K_{n,s}).
$$
\end{theorem}

\noindent\emph{Proof:}
Using the isomorphism $ L^2(\gamma \cup \I,\mathbb{R}^{N\times N})=L^2(\gamma,\mathbb{R}^{N\times N}) \oplus L^2(\I,\mathbb{R}^{N\times N})$ we can write $\tilde \K_{n,s}$ in matrix form as
$$\tilde \K_{n,s}=\le[\begin{array}{c|c}
0 & \KF_s \\
\hline
\KG & 0
\end{array}\ri],$$
where the operators $\KF_s:L^2(\gamma,\mathbb{R}^{N\times N}) \longrightarrow L^2(\I,\mathbb{R}^{N\times N})$ and $\KG: L^2(\I,\mathbb{R}^{N\times N}) \longrightarrow L^2(\gamma,\mathbb{R}^{N\times N})$ are defined respectively by the kernels
\be
	\bm\F_s(\lambda, z):=\frac{{\rm e}^{-\frac{z^2}2+2s(z-\lambda)-n \log(z)}}{2\pi i (\lambda- z)}\chi_\I(\lambda)\chi_{\gamma}(z),
	\label{Effe}\\
	\bm\G(z,w):=\frac{{\rm e}^{-\frac{z^2}2+w^2+n\log(w)}\bm\B_n(z)\bm\hB_n(w)}{2 \pi i (w-z)}\chi_{\gamma}(z)\chi_{\I}(w).\label{Gi}
\ee
Let us also introduce the Hilbert-Schmidt operator $\tilde{\K}'_{n,s}$ written in matrix form as

$$\tilde \K'_{n,s}:=\le[\begin{array}{c|c}
0 & -\KF_s \\
\hline
0 & 0
\end{array}\ri].$$
Through the identity
$$(\Id-\tilde \K_{n,s}')(\Id-\tilde \K_{n,s})= \le[\begin{array}{c|c}
\Id & \KF_s \\
\hline
0 & \Id
\end{array}\ri]\circ
\le[\begin{array}{c|c}
 \Id & -\KF_s \\
\hline
-\KG & \Id
\end{array}\ri] = \le[\begin{array}{c|c}
\Id-\KF_s\circ\KG  & 0  \\
\hline
-\KG & \Id
\end{array}\ri]$$
we get the following chain of equalities (see the Appendix B for the definition of $\mathrm{det}_2$):
\be
	\det(\Id-\tilde \K_{n,s})=\mathrm{det}_2(\Id-\tilde \K_{n,s})=\mathrm{det}_2(\Id-\tilde \K'_{n,s})\mathrm{det}_2(\Id-\tilde \K_{n,s})=\nonumber \\
	= \mathrm{det}_2(\Id-\KF_s\circ\KG){\rm e}^{-\mathrm{Tr}(\KF_s\circ\KG)}=\det(\Id-\KF_s \circ \KG)\label{eqdets}.
\ee
Now, using the formulas (\ref{Effe}) and (\ref{Gi}), we deduce that $\KF_s \circ \KG : L^2(\I,\mathbb{R}^{N\times N}) \longrightarrow L^2(\I,\mathbb{R}^{N\times N})$ has kernel explicitly given by the convolution
\be
	(\bm\F_s * \bm\G) (\lambda,w) =\oint_\gamma \frac{dz}{(2\pi i)^2}\frac{{\rm e}^{w^2-z^2+2 s(z-\lambda)+n\log(w)-n\log(z)}\bm\B_n(z) \bm\hB_n(w)}
	{(\lambda-z)(w-z)}.\label{compositionFG}
\ee
Finally we conjugate $(\KF_s \circ \KG)$ with the Fourier transform $\mathcal T$ such that
$$
(\mathcal T \bm F)(x)=\int_\I \frac{d\lambda}{\sqrt{\pi i}}{\rm e}^{-2\lambda x} \bm F(\lambda),\quad\quad (\mathcal T^{-1} \bm G)(\lambda)=\int_\bR \frac{d x}{\sqrt{\pi i}}{\rm e}^{2 \lambda x } \bm G(x).
$$
Then we obtain (using Cauchy's residue theorem) that the kernel $\bm\Cc(x,y)$ associated to $(\mathcal T\circ\KF_s \circ \KG\circ\mathcal T^{-1})$ is equal to
\bea
	 &\bm\Cc(x,y) = \ds\int_\I\frac{d\lambda}{\pi i}{\rm e}^{2\lambda(x-s)} \oint_\gamma dz\int_\I dw\frac{{\rm e}^{w^2-z^2+2sz-2wy+n\log(w)-n\log(z)}\bm\B_n(z) \bm\hB_n(w)}
	{(2\pi i)^2(\lambda-z)(w-z)}=\nonumber\\
	&\nonumber\\
	&=\left\{\begin{array}{c}
	\bm 0\quad\quad \mathrm{if}\; x<s\\
\\
	\ds\frac{2}{(2\pi i)^2}\int_{\mathcal{I}}dw\oint_{\gamma}dz \;
    \bm\B_n(z)\bm\hB_n(w)\ds\frac{{\rm e}^{w^2-2wy-z^2+2zx+n\log(w/z)}}{w-z} \quad \mathrm{if}\; x\geq s.\label{conjugation}
\end{array}\right.
\eea
and this latter, up to a conjugation with the operator of multiplication by ${\rm e}^{-\frac{x^2}2}$, is equal to $\bm K_n^\T(x,y)$. Hence \eqref{conjugation}, together with \eqref{compositionFG} and \eqref{eqdets}, gives \eqref{kernel}. \QED

We now introduce a Riemann-Hilbert problem related to the kernel $\tilde{\bm K}_{n,s}$ through the IIKS theory. In the following we denote $$\theta_{n}(\lambda,s):=\lambda^2-2\lambda s+n\log(\lambda).$$
\begin{problem}\label{HermiteRH}
	Find the sectionally analytic function $\bm\Gamma(\lambda)\in GL(N+p,\C)$ on $\C\backslash \{ \gamma \cup \I\}$ such that
	\be
		\left\{\begin{array}{ccc}
			\bm\Gamma_+(\lambda)=\bm\Gamma_-(\lambda)(\bm I-\bm G(\lambda)),\quad \lambda \in \gamma\cup\I ,\\
			\nonumber\\
			\bm\Gamma(\lambda)=\bm I+\ds\frac{\bm\Gamma_1}\lambda+\ds\frac{\bm\Gamma_2}{\lambda^2}+\cdots,\quad \lambda\rightarrow\infty.
		\end{array}\right.
	\ee
	with
	\be
	\bm G(\lambda) :=	\le[\begin{array}{ccccc}
					\bm 0 & {\rm e}^{\theta_{n}(\lambda,s)}\bm\hB_n^\T(\lambda) \\
					\bm 0 & \bm 0
					\end{array}\ri]\chi_{\I}(\lambda)+
				\le[ \begin{array}{ccccc}
					\bm 0 & \bm 0\\
					-{\rm e}^{-\theta_{n}(\lambda,s)} \bm\B_n^{\T}(\lambda) & \bm 0
					\end{array}\ri]\chi_{\gamma}(\lambda).
	\nonumber\ee
\end{problem}
\begin{theorem}
	The Fredholm determinant $\det(\Id-\chi_s\K_n)$ is equal to the isomonodromic tau function \eqref{omegamalgrange} related to the Riemann-Hilbert problem \ref{HermiteRH}. Hence, in particular, we have that
	\be\label{3.8}
		\partial_s\log\det(\Id-\chi_s\K_n) = \int_{\gamma\cup\I} \mathrm{Tr}\left(\bm\Gamma_-^{-1}(\lambda)(\partial_{\lambda}\bm\Gamma_-)(\lambda)\bm\Xi(\lambda)\right)\frac{d\lambda}{2\pi i},
	\ee
	where we denoted
	\be\label{3.9}
		\bm\Xi(\lambda) := \partial_s(\bm I-\bm G(\lambda))(\bm I-\bm G(\lambda))^{-1} = -\partial_s\bm G(\lambda)(\bm I+\bm G(\lambda)).
	\ee
\end{theorem}
\noindent\emph{Proof:}

Let's define the two $((N+p) \times N)$ matrices written block-wise as
\bes
	\vec{\bm f}(\lambda)&:=&\frac{1}{2\pi i} \le[\begin{array}{c}
					{\rm e}^{-2s\lambda}\bm I_N\chi_{\I}(\lambda) \\
					\\
					-{\rm e}^{-\frac{\lambda^2}2}\bm\B_n^\T(\lambda)\chi_{\gamma}(\lambda)
					\end{array} \ri],
\\\nonumber \\
\vec{\bm g}(\lambda)&:=&
  \le[\begin{array}{c}
		{\rm e}^{-\frac{\lambda^2}{2}+2s\lambda-n\log(\lambda)}\bm I_N\chi_\gamma(\lambda) \\ \\ {\rm e}^{\lambda^2 + n\log(\lambda)}\bm\hB_n(\lambda)\chi_\I(\lambda)  \end{array}\ri].
\ees
It is straightforward to verify that the following two equations are satisfied:
\be
	\tilde{\bm K}_{n,s}(\lambda,\mu)=\frac{\vec{\bm f}^\T(\lambda)\vec{\bm g}(\mu)}{\lambda-\mu},\quad
	\bm G(\lambda)= 2\pi i \vec{\bm f}(\lambda) \vec{\bm g}^\T(\lambda)\nonumber.
\ee
Hence, using Theorem \ref{thdetMalgrange} together with Theorem \ref{TH1}, we conclude that, denoting by $\tau_{JMU}$ the Jimbo-Miwa-Ueno tau function related to the Riemann-Hilbert problem \ref{HermiteRH}, we have
$$
\tau_{JMU}(s) = \det(\Id-\tilde\K_{n,s}) = \det(\Id-\chi_s\K_n),
$$
and, as a consequence, equation \eqref{3.8}. The second equality in the equation \eqref{3.9} comes from the fact that $\vec{\bm g}^\T(\lambda)\vec{\bm f}(\lambda) = \bm 0$.
\QED

\subsection{Non-commutative PIV}\label{SEC31}

We now specialize to the particular case in which $N=2$ and $\K_n$ is the Hermite-type CD kernel \eqref{kernel} associated with the MOP studied in Sections \ref{firstcase} and \ref{secondcase}.

In the first case we have $\bm\B_n : = z^{\bm J_2}\bm B_nz^{-\bm J_2}$ and $\bm\hB_n(w) := (\bm\B_n(w))^{-1}$, where $\bm B_n$ is the constant matrix \eqref{Bn1}. In the second case we have $\bm\B_n(z):= z^{2\bm J_2}\bm B_nz^{-\bm J_3}$ and $\bm\hB_n(w):=w^{\bm J_3}\hat{\bm B}_nw^{-2\bm J_2}$, where $\bm B_n$ and $\hat{\bm B}_n$ are defined in \eqref{Bn2}. Let us introduce, for these two different (sub)cases, the two matrices
\bes
	\bm T_{\bm A}(\lambda) &:=& \left[\begin{array}{cc}
					\frac{\theta_n(\lambda,s)}2\bm I_2-\bm J_2\log(\lambda) &\bm 0\\
					\bm 0 & -\frac{\theta_n(\lambda,s)}2\bm I_2-\bm J_2\log(\lambda)
					\end{array}\right]\quad\mathrm{and}\\
	\nonumber \\
	\bm T_{\bm B}(\lambda) &:=& \left[\begin{array}{cc}
					\frac{\theta_n(\lambda,s)}2\bm I_2-2\bm J_2\log(\lambda) & \bm 0\\
					\bm 0 & -\frac{\theta_n(\lambda)}2\bm I_3-\bm J_3\log(\lambda)
					\end{array}\right].
\ees
In the following we use the curly brackets to denote the anti-commutator between two matrices, i.e. $\{\bm x,\bm y\}:=\bm x\bm y+\bm y\bm x$. Moreover we denote with a prime the derivative with respect to $s$.
\begin{theorem}\label{theoncPIVfirstcase}
	Let $\bm B_n$ be as in \eqref{Bn1} and $\bm\Gamma(\lambda)$ be the solution of the Riemann-Hilbert problem \ref{HermiteRH} with jump
	\be\label{jumpfirstcase}
		\bm G(\lambda) =	\le[\begin{array}{ccccc}
					\bm 0 & {\rm e}^{\theta_{n}(\lambda,s)}\lambda^{-\bm J_2}\bm B_n^{-\T}\lambda^{\bm J_2} \\
					\\
					 \bm 0 & \bm 0
					\end{array}\ri]\chi_{\I}(\lambda)+
				\le[ \begin{array}{ccccc}
					\bm 0 & \bm 0\\
					\\
					-{\rm e}^{-\theta_{n}(\lambda,s)}\lambda^{-\bm J_2}\bm B_n^{\T}\lambda^{\bm J_2}  & \bm 0
					\end{array}\ri]\chi_{\gamma}(\lambda).
	\ee
Then
\be\label{traceformulafirstcase}
	\partial_s\log\det(\Id-\chi_s\K_n) = \mathrm{Tr}\Big((\bm \Gamma_1)_{22}-(\bm \Gamma_1)_{11}\Big),
\ee
where $\K_n$ is the integral operator with kernel $\bm K_n(x,y)$ given by \eqref{MVHKR}.\\

\noindent Moreover $\bm \Psi(\lambda):=\bm \Gamma(\lambda){\rm e}^{\bm T_{\bm A}(\lambda)}$ satisfies the Lax equations
$$
		\left\{ \begin{array}{ccc}
				\partial_\lambda \bm \Psi = \bm \A\bm \Psi =\le[\begin{array}{cc}
				(\lambda-s)\bm I_2+\left(\ds\left(\frac{n}2-z\right)\bm I_2-\bm J_2\right)\lambda^{-1} & \bm y-\ds\frac{\bm u\bm y}{2}\lambda^{-1}\\
				&\\
				2\bm y^{-1}\bm z+(\bm y^{-1}\bm z'-\bm y^{-1}\bm u\bm z)\lambda^{-1} & -(\lambda-s)\bm I_2-\left(\ds\frac{n}{2}\bm I_2+\bm J_2-\bm y^{-1}\bm z\bm y\right)\lambda^{-1}		
			\end{array}\right]\bm \Psi,&&\\
				\\
				\partial_s \bm \Psi = \bm \U\bm \Psi = \ds\le[\begin{array}{cc}
				-\lambda\bm I_2 & -\bm y\\
				&\\
				-2\bm y^{-1}\bm z & \lambda\bm I_2		
			\end{array}\right]\bm \Psi,&&
			\end{array}\right.
$$
where
\be\label{defvariablesfirstcase}
	\le\{\begin{array}{cc}
			\bm z:=-(\bm \Gamma_1)_{11}', \\
			\\
			\bm y:=-2(\bm \Gamma_1)_{12},\\
			\\
			\bm u:=(\bm \Gamma_1)_{12}'(\bm \Gamma_1)_{12}^{-1}+2s\bm I_2.
		\end{array}\right.
\ee
The compatibility conditions give the following coupled system of ODEs:
\be\label{zucouplefirstcase}
	\left\{\begin{array}{ccc}
		\bm u' &=& -\bm u^2+2s\bm u+4\bm z-2n\bm I_2 + \bm V_{\bm A},\\
		\\
		\bm z'' &=& 2\bm u'\bm z+2\bm u\bm z'-2s\bm z',
	\end{array}\right.
\ee
where $\bm V_{\bm A}:=2[\bm J_2,\bm y]\bm y^{-1}$.\\

\noindent Combining these two equations we obtain a non-commutative version of the derived PIV equation, in the form
\be\label{ncPIVfirstcase}
	\begin{array}{c}
	\bm u'''\!\!+[\bm u'',\bm u]\!-4(n+1+s^2)\bm u'\!\!-2\left(\lbrace \bm u',\bm u^2\rbrace+\bm u\bm u'\bm u\right)\\
\\
\hspace{3cm}+6s\lbrace \bm u', \bm u\rbrace +4\bm u(\bm u-s\bm I_2)+(\bm V'_{\bm A}-2(\bm u\bm V_{\bm A}))'+2s\bm V'_{\bm A}\! = \! \bm 0.\\
	\end{array}
\ee
\end{theorem}

In the second case we will be dealing with rectangular matrices. Hence in the following,  given an $(N\times p)$ rectangular matrix $\bm M$, with $N<p$ and linearly independent rows, we denote with $\bm M^\dagger$ the \emph{right inverse}\footnote{In fact, the definition of a right inverse of an $(N\times p)$ rectangular matrix $\bm M$ is not unique. For any invertible $(p\times p)$ matrix $\bm C$, a right inverse of $\bm M$ can be defined as $\bm M^\dagger := \bm C\bm M^\T(\bm M\bm C\bm M^\T)^{-1}$. Therefore we will have eventually a family of equations of the form \eqref{ncPIVsecondcase}. However we normalize all the computations assuming $\bm C=\bm I$.} of $\bm M$ defined as $\bm M^\dagger := \bm M^\T(\bm M\bm M^\T)^{-1}$.

\begin{theorem}\label{theoncPIVsecondcase}
	Let $\bm B_n,\hat{\bm B}_n$ be as in \eqref{Bn2} and $\bm \Gamma(\lambda)$ be the solution of the Riemann-Hilbert problem \ref{HermiteRH} with jump
$$
		\bm G(\lambda) =	\le[\begin{array}{ccccc}
					\bm 0 & {\rm e}^{\theta_{n}(\lambda,s)}\lambda^{-2\bm J_2}\hat{\bm B}_n^{\T}\lambda^{\bm J_3} \\
					\\
					 \bm 0 & \bm 0
					\end{array}\ri]\chi_{\I}(\lambda)+
				\le[ \begin{array}{ccccc}
					\bm 0 & \bm 0\\
					\\
					-{\rm e}^{-\theta_{n}(\lambda,s)}\lambda^{-2\bm J_2}\bm B_n^{\T}\lambda^{\bm J_3}  & \bm 0
					\end{array}\ri]\chi_{\gamma}(\lambda).
$$
Then
\be\label{traceformulasecondcase}
	\partial_s\log\det(\Id-\chi_s\K_n) = \mathrm{Tr}\Big((\bm \Gamma_1)_{22}-(\bm \Gamma_1)_{11}\Big),
\ee
where $\K_n$ is the integral operator with kernel $\bm K_n(x,y)$ given by \eqref{MVHKR2}.\\

\noindent Moreover $\bm \Psi(\lambda):=\bm \Gamma(\lambda){\rm e}^{\bm T_{\bm B}(\lambda)}$ satisfies the Lax equations
$$
		\left\{ \begin{array}{ccc}
				\partial_\lambda \bm \Psi = \bm \A\bm \Psi =\le[\begin{array}{cc}
				(\lambda-s)\bm I_2+\left(\ds\left(\frac{n}2-z\right)\bm I_2-2\bm J_2\right)\lambda^{-1} & \bm y-\ds\frac{\bm u\bm y}{2}\lambda^{-1}\\
				&\\
				2\bm y^{\dagger}\bm z+(\bm y^{\dagger}\bm z'-\bm y^{\dagger}\bm u\bm z)\lambda^{-1} & -(\lambda-s)\bm I_3-\left(\ds\frac{n}{2}\bm I_3+\bm J_3-\bm y^{\dagger}\bm z\bm y\right)\lambda^{-1}		
			\end{array}\right]\bm \Psi,&&\\
				\\
				\partial_s \bm \Psi = \bm \U\bm \Psi = \ds\le[\begin{array}{cc}
				-\lambda\bm I_2 & -\bm y\\
				&\\
				-2\bm y^{\dagger}\bm z & \lambda\bm I_3		
			\end{array}\right]\bm \Psi,&&
			\end{array}\right.
$$
where
$$
	\le\{\begin{array}{cc}
			\bm z:=-(\bm \Gamma_1)_{11}', \\
			\\
			\bm y:=-2(\bm \Gamma_1)_{12},\\
			\\
			\bm u:=(\bm \Gamma_1)_{12}'(\bm \Gamma_1)_{12}^{\dagger}+2s\bm I_2.
		\end{array}\right.
$$
The compatibility conditions give the following coupled system of ODEs:
$$
	\left\{\begin{array}{ccc}
		\bm u' &=& -\bm u^2+2s\bm u+4\bm z-2n\bm I_2 + \bm V_{\bm B},\\
		\\
		\bm z'' &=& 2\bm u'\bm z+2\bm u\bm z'-2s\bm z',
	\end{array}\right.
$$
where $\bm V_{\bm B}:= 4\bm J_2-2\bm y\bm J_3\bm y^\dagger$.\\

\noindent Combining these two equations we obtain a non-commutative version of the derived PIV equation, in the form
\be\label{ncPIVsecondcase}
	\begin{array}{c}
	\bm u'''\!\!+[\bm u'',\bm u]\!-4(n+1+s^2)\bm u'\!\!-2\left(\lbrace \bm u',\bm u^2\rbrace+\bm u\bm u'\bm u\right)\\
\\
\hspace{3cm}+6s\lbrace \bm u', \bm u\rbrace +4\bm u(\bm u-s\bm I_2)+(\bm V'_{\bm B}-2(\bm u\bm V_{\bm B}))'+2s\bm V'_{\bm B}\! = \! \bm 0.\\
	\end{array}
\ee
\end{theorem}

\begin{remark}

Before going into the proof let us remark that both the equations (\ref{ncPIVfirstcase}) and \eqref{ncPIVsecondcase} are equations for just one variable, namely $\bm y$; $\bm u$ and $\bm V$ (either $\bm V_{\bm A}$ or $\bm V_{\bm B}$) being functions of $\bm y$. The reason why we claim that this equation is a non-commutative version of the derived PIV is that, in both cases, if we assume that all the variables commute, we get the equation
$$
	u'''-4u'-6u^2u'+12u'u-4nu'+4u^2-4su-4s^2u' = 0,
$$
and the reader can easily verify that this latter equation is the derivative of the standard PIV equation
$$
	u'' = \frac{(u')^2}{2u}+\frac{3}2 u^3-4su^2+2(s^2+1+n)u-\frac{2n^2}{u}.
$$
\end{remark}

\noindent\emph{Proof of Theorem \ref{theoncPIVfirstcase}:}

Given $\bm \Gamma(\lambda)$ solution of the Riemann-Hilbert problem \ref{HermiteRH} with $\bm G(\lambda)$ as in \eqref{jumpfirstcase}, we have
\be\label{MarcoFormula}
	\int_{\gamma\cup\I}\mathrm{Tr}\left(\bm \Gamma_-^{-1}(\lambda)\partial_\lambda\bm \Gamma_-(\lambda)\bm \Xi(\lambda)\right)\frac{d\lambda}{2\pi i} = -	\res_{\lambda = \infty}\mathrm{Tr}\left(\bm \Gamma_-^{-1}(\lambda)\partial_\lambda\bm \Gamma_-(\lambda)\partial_s\bm T_{\bm A}\right),
\ee
where the (formal) residue above simply stands for minus the coefficient of the power $\lambda^{-1}$ in the asymptotic expansion of the argument. The formula \eqref{MarcoFormula} can be proven using Cauchy theorem and goes back to the article of Palmer \cite{Palmer:Zeros}. A very precise detailed derivation is given in \cite{BertolaIsoTau}, section 5.1.
Direct application of the formula \eqref{MarcoFormula} and the equation \eqref{3.8} yields the equation \eqref{traceformulafirstcase}.

Next we observe that $\bm \Psi(\lambda) = \bm \Gamma(\lambda){\rm e}^{\bm T_{\bm A}(\lambda)}$ solves a Riemann-Hilbert problem with constant jump, hence both $(\pa_\lambda\bm \Psi)\bm \Psi^{-1}$ and $(\pa_s\bm \Psi)\bm \Psi^{-1}$ are meromorphic functions on $\C^*$ so that, in particular, considering the singular behavior at $0$ at $\infty$, we get
\be
	\left\{\begin{array}{cc}
		&( \pa_\lambda \bm \Psi)\bm \Psi^{-1} = \bm \A = \lambda\bm \A_1+\bm \A_0+\lambda^{-1}\bm \A_{-1}\\
		&\nonumber\\
		& (\pa_s \bm \Psi)\bm \Psi^{-1} = \bm \U = \lambda\bm \U_1+\bm \U_0.
		\end{array}\right.
\ee
Let us start with $\bm \A$; in order to compute the coefficients $\bm \A_1,\bm \A_0,$ we compute the first terms of the asymptotic expansion at infinity of $(\pa_\lambda \bm \Psi)\bm \Psi^{-1}$ giving immediately\footnote{We denote with $\bm \sigma_3$ the standard Pauli matrix $\bm \sigma_3 = \mathrm{diag}(1,-1)$.}
$$
	\bm \A_1=\bm I_2\otimes\bm \sigma_3,\quad \bm \A_0=-s\bm I_2\otimes\bm \sigma_3+\left[\bm \Gamma_{1},\bm I_2\otimes\bm \sigma_3\right]=-s\bm I_2\otimes \bm \sigma_3+\le[\begin{array}{cc}
																								 \bm 0 & -2(\bm \Gamma_1)_{12}\\
																								 2(\bm \Gamma_1)_{21}  &\bm 0
																							       \end{array}\ri].
$$
The equation for $\bm \A_{-1}$ is slightly more complicated: we use again the expansion at infinity and we get
\be\label{2.17}
	\bm \A_{-1} = [\bm \Gamma_2,\bm I_2\otimes\bm \sigma_3] + [\bm I_2\otimes\bm \sigma_3,\bm \Gamma_1]\bm \Gamma_1+\frac{n}2 \bm I_2\otimes\bm \sigma_3-\bm J_2\otimes\bm I_2-[\bm \Gamma_1,s\bm I_2\otimes\bm \sigma_3].
\ee

In order to simplify the expression above we use the fact that the $\lambda^{-1}$-term of $(\pa_s\bm \Psi)\bm \Psi_{-1}$ term is identically zero, giving
\be\label{2.18}
	\bm \Gamma_1' = [\bm \Gamma_2,\bm I_2\otimes\bm \sigma_3] + [\bm I_2\otimes\bm \sigma_3,\bm \Gamma_1]\bm \Gamma_1,
\ee
leading eventually to
$$
	\bm \A_{-1} = \bm \Gamma_1' - [\bm \Gamma_1,s\bm I_2\otimes\bm \sigma_3] + \frac{n}2 \bm I_2\otimes\bm \sigma_3 - \bm J_2\otimes\bm I_2 .
$$
The block-diagonal part of the equation above gives
\be\label{2.20}
	(\bm \Gamma_1)'_{11} = 2(\bm \Gamma_1)_{12}(\bm \Gamma_1)_{21},\quad (\bm \Gamma_1)'_{22} = -2(\bm \Gamma_1)_{21}(\bm \Gamma_1)_{12},
\ee
and so, combining (\ref{2.17}), (\ref{2.18}) with the definitions (\ref{defvariablesfirstcase}) and using the relations \eqref{2.20} we get
$$
	\bm \A_0 = \le[\begin{array}{cc}
					-s\bm I_2 & \bm y\\
					&\\
					2\bm y^{-1}\bm z & s\bm I_2
			\end{array}\ri],\quad
	\bm \A_{-1} = \le[\begin{array}{cc}
					-\bm z+\ds\frac{n}2\bm I_2-\bm J_2 & -\ds\frac{\bm u\bm y}2\\
					&\\
					(\bm \A_{-1})_{21} & \bm y^{-1}\bm z\bm y-\ds\frac{n}{2}\bm I_2-\bm J_2
			\end{array}\ri].
$$
In order to get the missing term $(\bm \A_{-1})_{21}$ we use the Lax equation
$$
	\pa_s\bm \A-\pa_\lambda\bm \U = [\bm \U,\bm \A];
$$
the $(1,1)$ term giving
$$
(\bm \A_{-1})_{21} = \bm y^{-1}\bm z'-\bm y^{-1}\bm u\bm z.
$$
In this way we arrive to the stated expression for $\bm \A$. The computation for $\bm \U$, since we have no singularities at the origin, is simpler and the asymptotic at infinity of $(\pa_s \bm \Psi)\bm \Psi^{-1}$ gives immediately
$$
	\bm \U_1 = -\bm I_2\otimes\bm \sigma_3,\quad \bm \U_0 = \le[\begin{array}{cc}
					\bm 0 & 2(\bm \Gamma_1)_{12}\\
					&\\
					-2(\bm \Gamma_1)_{21} &  \bm 0
			\end{array}\ri] =
			\le[\begin{array}{cc}
					\bm 0 & -\bm y\\
					&\\
					-2\bm y^{-1}\bm z & \bm 0
			\end{array}\ri].
$$

The first equation in \eqref{zucouplefirstcase} is just the $(1,2)$ (block) entry of the equation. For the second one we start observing that the off-diagonal terms of the relation (\ref{2.18}) gives the equations
\be\label{offdiag}
	(\bm \Gamma_1)'_{12} = -2(\bm \Gamma_2)_{12} + 2(\bm \Gamma_1)_{12}(\bm \Gamma_1)_{22},\quad (\bm \Gamma_1)'_{21} = 2(\bm \Gamma_2)_{21} - 2(\bm \Gamma_1)_{21}(\bm \Gamma_1)_{11}.
\ee
Then we go on analyzing the asymptotic expansion at infinity of $(\partial_\lambda \bm \Psi)\bm \Psi^{-1}$; the $\lambda^{-2}$-term (which is identically zero) gives the equation
\be\label{lunghissima}
\begin{array}{c}
\bm \Gamma_1\! =\! [\bm \Gamma_3,\bm I_2\otimes\bm \sigma_3]\!+\![s\bm I_2\otimes\bm \sigma_3,\bm \Gamma_2]\!+\!\left[\bm \Gamma_1,\displaystyle\frac{n}2\bm I_2\otimes\bm \sigma_3-\bm J_2\otimes\bm I_2\right]\!\\
\\
\qquad\qquad\qquad\qquad\qquad+[\bm\Gamma_1,s\bm I_2\otimes\bm\sigma_3]\bm\Gamma_1\!+\![\bm\Gamma_1,\bm I_2\otimes\bm\sigma_3]\bm\Gamma_1^2\!+\![\bm I_2\otimes\bm\sigma_3,\bm\Gamma_1]\bm\Gamma_2\!+\![\bm I_2\otimes\bm\sigma_3,\bm\Gamma_2]\bm\Gamma_1.\\
\end{array}
\ee
The $(1,1)$-entry of this equation gives
\be\label{penultima}
\begin{array}{c}
(\bm \Gamma_1)_{11}=-2s(\bm \Gamma_1)_{12}(\bm \Gamma_1)_{21}-2(\bm \Gamma_1)_{12}(\bm \Gamma_1)_{21}(\bm \Gamma_1)_{11}\\
\\
\hspace{5cm}-2(\bm \Gamma_1)_{12}(\bm \Gamma_1)_{22}(\bm \Gamma_1)_{21}+2(\bm \Gamma_1)_{12}(\bm \Gamma_2)_{21}+2(\bm \Gamma_2)_{12}(\bm \Gamma_1)_{21},\\
\end{array}
\ee
and combining (\ref{offdiag}) with (\ref{penultima}) we get the second equation in \eqref{zucouplefirstcase} (the computation is lengthy but completely straightforward). The equation \eqref{ncPIVfirstcase} is obtained simply expressing $\bm z$ in function of $\bm u$ using the first equation in \eqref{zucouplefirstcase} and then substituting in the second one.\QED

Since the proof of the Theorem \ref{theoncPIVsecondcase} is formally identical to the one of Theorem  \ref{theoncPIVfirstcase} we will not write it here. One has just to substitute $\bm J_2$ with $\bm J_3$ where appropriate, and substitute $\bm y^{-1}$ with $\bm y^\dagger$. Indeed, going through the computations of the proof above, is easy to verify that we use just the property $\bm y\bm y^{-1} = \bm I_2$, while $\bm y^{-1}\bm y = \bm I_2$ is never used.

\subsubsection{A symmetric formulation of PIV}\label{3.1.1}

While formulas \eqref{ncPIVfirstcase},\eqref{ncPIVsecondcase} establish a direct connection with the classical PIV equation and Tracy-Widom results \cite{TWPIV}, in view of equations \eqref{traceformulafirstcase} and \eqref{traceformulasecondcase} it would be desirable to write a system of ODEs in which both the entries $(\bm \Gamma_1)_{11}$ and $(\bm \Gamma_1)_{22}$ appear explicitly as dependent variables. To this aim the symmetric formulation of PIV given (in the scalar case) by Aratyn, Gomes and Zimerman in \cite{AGZ1,AGZ2} is particularly suitable. A similar description can be given in our non-commutative case. We introduce new notations for the entries of the Lax matrices so to match with the cited articles.

\begin{theorem}
	Let $\bm \Psi(\lambda) := \bm \Gamma(\lambda){\rm e}^{\bm T_{\bm A}(\lambda)}$ be as in Theorem \ref{theoncPIVfirstcase} and denote
	\bes
	\left\{\begin{array}{ccc}
		\bm q &:=& 2(\bm \Gamma_1)_{12},\\
		\\
		\bm r &:=& -2(\bm \Gamma_1)_{21},\\
		\\
		\bm \rho_R &:=& 4(\bm \Gamma_1)_{11},\\
		\\
		\bm \rho_L &:=& -4(\bm \Gamma_1)_{22}.
	\end{array}\right.
	\ees
Then the related Lax matrices $\bm \A$ and $\bm \U$ reads
\bes
	\bm \U = \left[\begin{array}{cc}
									-\lambda\bm I_2 & \bm q\\
									\bm r & \lambda\bm I_2
									\end{array}\right],
\ees
\bes
	\bm \A = \left[\begin{array}{cc}
									(\lambda-s)\bm I_2 & -\bm q\\
									-\bm r & -(\lambda-s)\bm I_2
									\end{array}\right] + \frac{1}{4\lambda}\left[\begin{array}{cc}
										\bm \rho'_R + 2n\bm I_2 -4\bm J_2 & 4 s \bm q + 2 \bm q'\\
										4 s \bm r - 2 \bm r' &  -\bm \rho'_L- 2n\bm I_2 - 4\bm J_2
									\end{array}\right],
\ees
while Lax equations read
\be\label{symncPIVfirstcase}
\left\{\begin{array}{cc}
	\bm \rho_R = 2s\bm q\bm r + \bm q'\bm r-\bm q\bm r',\\
	\\
	\bm \rho_L = 2s\bm r\bm q + \bm r\bm q'-\bm r'\bm q,\\
	\\
	-s\bm q'+\ds\frac{1}{2}(-\bm q''+2\bm q\bm r\bm q) = (1+2n)\bm q + 2[\bm q,\bm J_2],\\
	\\
	s\bm r' -\ds\frac{1}{2}(\bm r''-2\bm r\bm q\bm r) = (-1+2n)\bm r + 2[\bm r,\bm J_2].
\end{array}
\right.
\ee
\end{theorem}
\begin{theorem}
	Let $\bm \Psi(\lambda) := \bm \Gamma(\lambda){\rm e}^{\bm T_{\bm B}(\lambda)}$ be as in Theorem \ref{theoncPIVsecondcase} and denote
	\bes
	\left\{\begin{array}{ccc}
		\bm q &:=& 2(\bm \Gamma_1)_{12},\\
		\\
		\bm r &:=& -2(\bm \Gamma_1)_{21},\\
		\\
		\bm \rho_R &:=& 4(\bm \Gamma_1)_{11},\\
		\\
		\bm \rho_L &:=& -4(\bm \Gamma_1)_{22}.
	\end{array}\right.
	\ees
Then the related Lax matrices $\bm \A$ and $\bm \U$ reads
\bes
	\bm \U = \left[\begin{array}{cc}
									-\lambda\bm I_2 & \bm q\\
									\bm r & \lambda\bm I_3
									\end{array}\right],
\ees
\bes
	\bm \A = \left[\begin{array}{cc}
									(\lambda-s)\bm I_2 & -\bm q\\
									-\bm r & -(\lambda-s)\bm I_3
									\end{array}\right] + \frac{1}{4\lambda}\left[\begin{array}{cc}
										\bm \rho'_R + 2n\bm I_2 -8\bm J_2 & 4 s \bm q + 2 \bm q'\\
										4 s \bm r - 2 \bm r' &  -\bm \rho'_L- 2n\bm I_3 -4\bm J_3
									\end{array}\right],
\ees
while Lax equations read
\be\label{symncPIVsecondcase}
\left\{\begin{array}{cc}
	\bm \rho_R = 2s\bm q\bm r +\bm q'\bm r-\bm q\bm r',\\
	\\
	\bm \rho_L = 2s\bm r\bm q + \bm r\bm q'-\bm r'\bm q,\\
	\\
	-s\bm q'+\ds\frac{1}{2}(-\bm q''+2\bm q\bm r\bm q) = (1+2n)\bm q + 4[\bm q,\bm J_2],\\
	\\
	s\bm r' -\ds\frac{1}{2}(\bm r''-2\bm r\bm q\bm r) = (-1+2n)\bm r + 2[\bm r,\bm J_3].
\end{array}
\right.
\ee
\end{theorem}
\begin{remark}
The system of equations above \eqref{symncPIVfirstcase} and \eqref{symncPIVsecondcase} are the non-commutative analogues of equations (2.4) in \cite{AGZ1}, which are equivalent to the sigma-form of PIV (see equations (2.6),(2.7) in \cite{AGZ1})
\end{remark}

\noindent\emph{Proof:}
Both theorems are proven in the same way and the proof consists, essentially, on rewriting the same equations as in Theorems \ref{theoncPIVfirstcase} and \ref{theoncPIVsecondcase} with different variables.\\
In particular the third and the fourth equations in \eqref{symncPIVfirstcase},\eqref{symncPIVsecondcase} come from the the Lax equation
$$
\pa_s\bm \A-\partial_\lambda\bm \U = [\bm \U,\bm \A],
$$
together with the already used fact that $\bm \rho'_R = -\bm q\bm r$ and $\bm \rho'_L = -\bm r\bm q$ (see equations \eqref{2.20}).
The other two equations come from the (already used) equations (\ref{offdiag}) combined with (\ref{penultima}) and
$$
(\bm \Gamma_1)_{22} = 2\left(s(\bm \Gamma_1)_{21}(\bm \Gamma_1)_{12} + (\bm \Gamma_1)_{21}(\bm \Gamma_1)_{11}(\bm \Gamma_1)_{12}+(\bm \Gamma_1)_{21}(\bm \Gamma_1)_{12}(\bm \Gamma_1)_{22}-(\bm \Gamma_1)_{21}(\bm \Gamma_2)_{12}-(\bm \Gamma_2)_{21}(\bm \Gamma_1)_{12}\right),
$$
this latter coming from the $(2,2)$-entry of (\ref{lunghissima}). \QED

\section{Concluding remarks}\label{AP0}

In this paper we have shown (for the first time, to the best of our knowledge) some integral representations of two particular examples of Hermite-type MOP on the real line, and then deduced a double integral representation of the related Christoffel-Darboux kernel $\bm K_n(x,y)$. This was the starting point for the study of the Fredholm determinant $\det(\Id-\chi_s\K_n)$, where $\K_n$ is the integral operator with kernel $\bm K_n(x,y)$. Using some Riemann-Hilbert techniques, we related this Fredholm determinant with a non-commutative version of the Painlev\'e IV equation. We remark that the type of orthogonality we started from is essential in order to get that specific equation. In complete analogy with the scalar case \cite{TWPIV}, for instance, we expect that Laguerre-type matrix orthogonal polynomials should be related to some non-commutative version of the Painlev\'e V equation. We plan to investigate on this issue in subsequent works.

Another interesting question that arises is how these kernels $\bm K_n(x,y)$ behave if we consider scaling limit as $n\to\infty$. It is very well known that, after rescaling appropriately the variables, the Hermite kernel converges to the Airy kernel in the following way
$$
\lim_{n\to\infty}\frac{1}{\sqrt{2}n^{1/6}}K_n^{\footnotesize{\textrm{Hermite}}}\left(\sqrt{2 n}+\frac{x}{\sqrt{2}n^{1/6}},\sqrt{2 n}+\frac{y}{\sqrt{2}n^{1/6}}\right)=K_{\footnotesize{\textrm{Ai}}}(x,y),
$$
where $x$ and $y$ are in a bounded set (see \cite{AFvM} for a very nice and elementary deduction of this convergence, even in a more general setting than what needed here).

In the matrix case, for the two Hermite-type kernels we study in this paper (see \eqref{MVHK1c} and \eqref{MVHK2c}) there will be a \emph{scalar} behavior as $n\to\infty$, i.e.
$$
\lim_{n\to\infty}\frac{1}{\sqrt{2}n^{1/6}}\bm K_n\left(\sqrt{2 n}+\frac{x}{\sqrt{2}n^{1/6}},\sqrt{2 n}+\frac{y}{\sqrt{2}n^{1/6}}\right)=K_{\Ai}(x,y)\bm I_N.
$$
In particular, this property holds for the examples \eqref{MVHKR} and \eqref{MVHKR2} ($N=2$). This is easy to see from the integral representation  \eqref{MVHK1c} (analogous for \eqref{MVHK2c}) using classical steepest descent methods, as in \cite{AFvM}. Indeed the ``matrix part'' of the kernel can be written as $z^{\bm J_N} \bm B_n\left(\frac{w}{z}\right)^{\bm J_N}\bm B_n^{-1} w^{-\bm J_N}$, but under the given rescaling\footnote{Recall that the variables $w$ and $z$ are rescaled as $w=\sqrt{\frac{n}{2}}\left(1+\frac{\mu}{n^{1/3}}\right)$ and $z=\sqrt{\frac{n}{2}}\left(1+\frac{\lambda}{n^{1/3}}\right)$, see \cite{AFvM}.} $\frac{w}{z}\to1$ as $n\to\infty$, therefore the matrix part converges to the identity matrix $\bm I_N$ as $n\to\infty$, no matter the choice of $\bm B_n$.

The Airy kernel is related, as it is well known, to the Painlev\'e II equation \cite{TWAiry}, and this relation extends to the matrix case for a specific type of \emph{matrix} Airy kernel \cite{BC2}.
For the two Hermite-type polynomials we study in this paper, the scaling limit of the corresponding CD kernel has a scalar behavior, so that the non-commutativity disappears and the corresponding matrix Painlev\'e II equation is nothing but $N$ non--interacting copies of the scalar equation. It would be interesting to find other Hermite-type MOP whose kernels do not behave in this way, something that in principle is not an easy task. This consideration, however, goes beyond the scope of this paper, and will be pursued elsewhere.

\medskip

\noindent{\bf Acknowledgements:}

The first author is grateful to V. Retakh and V. Rubtsov for many interesting discussions about their works on non-commutative integrable systems.

\appendix
\renewcommand{\theequation}{\Alph{section}.\arabic{equation}}

\section{A brief reminder of regularized Fredholm Determinants}\label{AP1}
We refer to \cite{Si} for the relevant details: we shall need only the elementary facts which we recall here. In general the Fredholm determinant of an operator of the form $(\Id - G)$ can be defined only when $G$ is of trace class. Recall that if $G$ is represented as an integral operator on a (separable) Hilbert  $L^2(X,\d\mu)$ with kernel $G(x,y)$ (we abuse notation here) then
\be
\det(\Id  - G) = 1 + \sum_{n=1}^\infty \frac 1 {n!} \int_{X^n} \det [G(x_i,x_j)]_{i,j\leq n}\prod_{i=1}^n \d\mu(x_i).
\label{FredhExp}\ee
There are other trace ideals $\mathcal I_p$, $p\in \mathbb N,$ which means that $G^p$ is trace-class \cite{Si}; in particular $\mathcal I_2$ consists of Hilbert-Schmidt operators.   For $G\in \mathcal I_p$ one can define following Carleman a regularized determinant $\det_p(\Id -G)$ which has the same main property of vanishing iff the operator is not invertible. In particular for Hilbert-Schmidt operators one has
\bes
\mathrm{det}_2(\Id - G):= 1 + \sum_{n=1}^\infty \frac 1 {n!} \int_{X^n} \det [G(x_i,x_j) (1-\delta_{ij})]_{i,j\leq n}\prod_{i=1}^n \d\mu(x_i),
\ees
that is, one simply omits the diagonal elements in the determinants under the integral sign. This determinant has the properties
\begin{itemize}
\item if $G$ is also trace-class then
\bes
\mathrm{det}_2(\Id - G) = \det (\Id - G) {\rm e}^{-{\rm tr} G}.
\ees
\item
if $G_1,G_2$ are Hilbert-Schmidt operators (and hence $G_1G_2$ is trace class) then
\bes
\mathrm{det}_2(\Id - G_1) \mathrm{det}_2 (\Id-G_2) = \mathrm{det}_2(\Id - G_1 - G_2 + G_1G_2) {\rm e}^{-{\rm tr} (G_1G_2)}.
\ees
\end{itemize}
An interesting occurrence (which is used in this article) is that if $G$ is just Hilbert-Schmidt but its kernel vanishes on the diagonal $G(x,x)\equiv 0$ then the series defining $\det_2(\Id-G)$ is identical to the regular $\det(\Id- G)$. The reason for still wanting to distinguish $\det_2$ from $\det$ in this case is simply that $G$ may fail to have a trace and in a different basis the ordinary $\det$ may simply be ill-defined.

\section{Integrable kernels and isomonodromic tau functions}\label{AP2}

In this appendix we recall some basic facts about integrable kernels \`a la Its-Izergin-Korepin-Slavnov \cite{IIKS} and their connections with isomonodromic tau functions, recalling in particular a theorem proved in \cite{BC1} (see also \cite{BC2}).
Given a piecewise smooth oriented curve ${\cal C}$ on the complex plane (possibly extending to infinity) and two matrix-valued functions
$$
\f,\g:\mathcal C\longrightarrow {\mathrm{Mat}_{p\times k}(\C)},
$$
we define the kernel $\bm K$ as
$$
\bm K(\lambda,\mu):=\frac{\f^{\mathrm T}(\lambda)\g(\mu)}{\lambda-\mu}.
$$
We say that such kernel is integrable if $\f^{\mathrm T}(\lambda)\g(\lambda)=\bm 0$ (so that it is non-singular on the diagonal). We are interested in the operator $K:L^2({\cal C},\C^k)\rightarrow L^2({\cal C},\C^k)$ acting on $k$-vector functions via the formula
$$
(K\bm h)(\lambda)=\int_{\cal C}\bm K(\lambda,\mu)\bm h(\mu)d\mu,
$$
and, in particular, we are interested in the Fredholm determinant $\det(\Id-K)$ defined as in (\ref{FredhExp}).
The key observation is that, denoting with $\partial$ the differentiation with respect to any auxiliary parameter on which $K$ may depend, we obtain the formula
\be
	\partial \log\det(\Id-K)=-\mathrm{Tr}((\Id+R)\partial K),
\label{resolvent}\ee
where $R$ is the resolvent operator, defined as $R=(\Id-K)^{-1}K$. Moreover $R$ is again an integrable operator, i.e.
$$
R(\lambda,\mu)=\frac{\bm F^{\mathrm T}(\lambda)\bm G(\mu)}{\lambda-\mu},
$$ and $\bm F,\bm G$ can be found solving the following RH problem:
\be\label{appendixRH}\left\{\begin{array}{ccc}
		\bm \Gamma_+(\lambda)&=&\bm \Gamma_-(\lambda)\bm M(\lambda),\quad\lambda\in{\cal C},\\
		\\
		\bm \Gamma(\lambda)& =&\bm I+\mathcal O(\lambda^{-1}),\quad\lambda\longrightarrow\infty,\\
		\\
	\bm M(\lambda)& =&\bm I-2\pi i\f(\lambda)\g^{\mathrm T}(\lambda).
	\end{array}\right.
\ee
More precisely we have the two equalities
\bes
  \bm F(\lambda)=\bm \Gamma(\lambda)\bm f(\lambda),\qquad
		\bm G(\lambda) =(\bm \Gamma^{-1})^{\mathrm T}(\lambda)\bm g(\lambda).
\ees
Now suppose, as in the formula \eqref{resolvent}, that the operator $K$ (and hence the Riemann-Hilbert problem \eqref{appendixRH} depends smoothly on a certain parameter set of parameters\footnote{In the case treated in this article, we just have one parameter, namely $s$.}. On the space of these deformation parameters, we introduce the following one-form (here below $\partial$ denotes a vector in the space of deformation parameters)
\bea
	\omega_M(\partial)&\&:= \int_{\cal C}\mathrm{Tr}\Big(\bm \Gamma_-^{-1}(\lambda)\partial_\lambda\bm \Gamma_-(\lambda)\bm \Xi_\partial(\lambda )\Big)\frac{\d \lambda}{2\pi i},
	\label{omegamalgrange}\\
	\nonumber\bm \Xi_\partial(\lambda)&\& :=\partial \bm M(\lambda)\bm M^{-1}(\lambda)
\eea
The definition (\ref{omegamalgrange}) is posed for arbitrary jump matrices; in the case of the Riemann-Hilbert problem \eqref{appendixRH} the spontaneous question arises as to whether $\omega_M$ in (\ref{omegamalgrange}) and the Fredholm determinant are related. The answer is positive  within a certain explicit correction term, as in the theorem below

\begin{theorem}[\cite{BC1}]\label{thdetMalgrange}\footnote{Actually the article \cite{BC1} treats the case $k=1$, but the proof does not change considering this more general case.}
Let $\f(\lambda;\vec s), \g(\lambda;\vec s): \mathcal{C} \times S \longrightarrow {\mathrm{Mat}_{p\times k}(\mathbb{C})}$ and consider the Riemann-Hilbert problem with jumps as in \eqref{appendixRH}. Given any vector field $\partial$ in the space of the parameters $S$ of the integrable kernel we have the equality
\bes
		\omega_M(\partial)=\partial\log\det(\Id-K)+H(\bm M),
\ees
	where $\omega_M(\partial)$ is as in (\ref{omegamalgrange}) and
$$
H(\bm M):=H_1(\bm M)-H_2(\bm M)=\int_{\cal C} \mathrm{Tr}\Big(\partial\f'^\T \g+\f'^\T\partial\g \Big)d\lambda -2\pi i\int_{\cal C}\mathrm{Tr}( \g^\T \f'
\partial\g^\T\f) d\lambda.
$$
\end{theorem}
In the cases we treat in the article, moreover, we have $H(\bm M)=0$.
Hence it is possible to define, up to normalization, the isomonodromic tau function $\tau_{JMU}:=\exp(\int\omega_M)$ and this object, thanks to the previous theorem, will coincide with the Fredholm determinant $\det(\Id-K)$.

\end{document}